\documentclass[10pt]{article}
\expandafter\let\csname equation*\endcsname\relax
\expandafter\let\csname endequation*\endcsname\relax
\usepackage{amsmath}
\usepackage{amssymb}
\usepackage{graphicx}
\usepackage{amsfonts}
\usepackage{xcolor}
\usepackage{comment}

\usepackage{style}
\usepackage{algorithm,algpseudocode}

\DeclareMathOperator{\ve}{\varepsilon}

\makeatletter
\newcommand{\mainmatter}{%
  \setcounter{footnote}{0}%
  \patchcmd{\@makefntext}{\fnsymbol}{\arabic}{}{}%
  \patchcmd{\@thefnmark}{\fnsymbol}{\arabic}{}{}%
  \def\@makefnmark{\textsuperscript{\arabic{footnote}}}%
}
\makeatother

\title{Memorization and Generalization in Generative Diffusion under the Manifold Hypothesis}

\author[1]{Beatrice Achilli%
}
\author[3]{Luca Ambrogioni%
}
\author[1,2]{Carlo Lucibello%
}
\author[1,2]{Marc M\'ezard%
}
\author[1,2]{Enrico Ventura%
}
\affil[1]{Department of Computing Sciences, 
Bocconi University, Milan, Italy.}
\affil[2]{Bocconi Institute for Data Science and Analytics (BIDSA), Milan, Italy.}
\affil[3]{Donders Institute,
Radboud University,
Nijmegen, The Netherlands.}
\date{\vspace{-8ex}}

\begin{document}

\maketitle



\vspace{10pt}
\begin{abstract}

We study the memorization and generalization capabilities of a Diffusion Model (DM) in the case of structured data defined on a latent manifold. 
We specifically consider a set of $P$ data points in $N$ dimensions lying on a latent subspace of dimension $D = \alpha_D N$, according to the Hidden Manifold Model (HMM). 
Our analysis considers a reverse process given by the empirical score function as a proxy of the true one, and then precisely characterizes the process in the high-dimensional limit in which $P=\exp(\alpha N)$ and $N$ is large, by exploiting a connection with the Random Energy Model (REM). We provide evidence for the existence of an onset time, $t_{o}$, when traps appear in the time-varying potential, although they do not affect typical trajectories. The size of the basins of attraction of such traps is computed at any time. 
Moreover, we derive the collapse time, $t_{c} < t_{o}$, at which trajectories fall in the basin of one of the training points, implying memorization. 
An explicit formula for $t_c$ is given as a function of $P$ and the ratio $\alpha_D$, proving that the curse of dimensionality issue does not hold for highly structured data, i.e. $\alpha_D\ll 1$, regardless of the non-linearity of the manifold surface. We also prove that collapse coincides with the condensation transition in the REM. 
Finally, the degree of generalization of DMs is formulated in terms of the Kullback-Leibler divergence between the exact distribution and the one obtained at time $t$ of the reverse process. We show the existence of an additional time $t_{g}<t_{c}<t_{o}$ such that the distance between the reverse distribution and the ground-truth is minimal. Counter-intuitively, the best generalization performance is found within the memorization phase of the model. We conclude that the generalization performance of DMs benefit from highly structured data since $t_g$ approaches zero faster than $t_c$ when $\alpha_D \rightarrow 0$.  

\end{abstract}

%
%
%
%
%
\makeatletter
\def\@mkboth#1#2{}
\newlength\appendixwidth
\preto\appendix{\addtocontents{toc}{\protect\patchl@section}}
\newcommand{\patchl@section}{%
  \settowidth{\appendixwidth}{\textbf{Appendix }}%
  \addtolength{\appendixwidth}{1.5em}%
  \patchcmd{\l@section}{1.5em}{\appendixwidth}{}{\ddt}%
}
\makeatother

\section{Introduction}
Generative diffusion models \citep{sohldickstein2015deep} have reached the state-of-the-art performance on image \citep{ho2020denoising, song2021score}, sound \citep{chen2020wave} and video generation \citep{ho2022videodiffusionmodels} by synthesizing data through a denoising process that can be expressed as a stochastic differential equation \citep{song2021score}. Recent work has established deep connections between the framework of generative diffusion and well-known phenomena in statistical physics \citep{Biroli_2023, montanari_posterior_2023, montanari_sampling_2023, ambrogioni2024thermo}. As an example, it was shown that class separation during the generative dynamics of diffusion models can be described in terms of symmetry breaking phase transitions \citep{raya_spontaneous_2023, Biroli_2023,biroli_dynamical_2024}, which are the result of a Curie-Weiss self-consistency condition implicit in the fixed-point structure of the score function \citep{ambrogioni2024thermo}. The presence of hierarchically organized and semantically meaningful phase transitions was also demonstrated in \citep{sclocchi2024phasetransitiondiffusionmodels}. 
Furthermore, it was recently shown that the generative dynamics of diffusion models is closely related to the retrieval dynamics of modern Hopfield networks \citep{ambrogioni_search_2023, hoover_memory_2023}, which are a class of associative memory models with exponential capacity \citep{krotov2016dense, demircigil2017model, krotov2023new, ramsauer2020hopfield}. In Ref. \cite{lucibello_exponential_2023}, the authors have shown how to compute the exponential rate for the capacity using the formalism of the Random Energy Model (REM) from spin glass theory \cite{Derrida1981}. 

Closet to our work, in Ref. \cite{biroli_dynamical_2024} the REM formalism
is used to characterize the memorization phenomenon in diffusion models in a way that mirrors the study of memory capacity of Hopfield models. These techniques were also used in \citep{achilli2024losing} to characterize the closure of gaps in the spectrum of the Jacobian of the score corresponding to \emph{geometric memorization} effects, where sub-spaces of the target distributions are lost due to fine sample size. 

In this paper, we provide a detailed theoretical analysis of generative diffusion models when the data are sampled from a low-dimensional, possibly non-linear, extending previous work that makes use of the REM technique. The paper is organized as follows:

\begin{itemize}
    \item In Section \ref{sec:hmm} we introduce the Hidden Manifold Model (HMM), which will serve as our data-generating model throughout the paper.
    \item Section \ref{sec:diffusion} provides background on Diffusion Models (DMs) and
    the two types of score functions we consider, the true one and the empirical one. 
    \item Section \ref{sec:REM} reviews the Random Energy Model (REM) formalism. This will be the workhorse of our analysis of DMs.
    \item In Section \ref{sec:memo}, we analyze the way DMs memorize data living on a manifold.  
    When using the empirical score function as an approximation of the true one, we highlight the presence of two dynamical phase transitions when simulating the reverse process with time $t$ going from $+\infty$ to $0$. 
    \begin{enumerate}
        \item The first one, at time $t_o$, is called the onset transition. It is when basins of attraction arise in correspondence of most data points, but they are not large enough to affect typical trajectories.
        \item The second, at time $t_c < t_o$, is called collapse transition \cite{biroli_dynamical_2024}. It corresponds to typical diffused particles being trapped in the potential well of one of the data points, with no chance of escaping it for the rest of the evolution. These last results are consistent with the recent analysis of Ref. \cite{macris}.  
    \end{enumerate} 
    For generically distributed data points, we show that the collapse transition corresponds to the condensation transition in the REM. Moreover, we show that for $t > t_c$ the empirical score is close to the true score.
    \item Finally, in Section \ref{sec:generalization} we analyze the problem of generalization in DMs  driven by the empirical score, using two approaches. We first compute the optimal stopping time $t_g$ which is the time at which the KL divergence between the diffused empirical distribution and the target distribution is minimal. We use
 the REM formalism again to compute this stopping time $t_g$. We find that it is always located in the condensed phase, i.e. $t_{g} < t_c$, a phenomenon that as been observed recently in the related framework of kernel approximations to large dimensional densities \cite{biroli_kernel_2024}. The optimal time $t_g$ is found to shrink to zero values faster than $t_c$ when the latent dimension of the data decreases. 
 In a second approach, we combine results obtained via REM formalism with random matrix computations (as performed in Ref. \cite{ventura2024spectral}), in order to deduce an empirical generalization criterion for DMs sampling before memorization.  
    
\end{itemize}

\section{Modeling the Manifold Hypothesis}
\label{sec:hmm}
In this paper we focus on data points generated by a Hidden Manifold Model (HMM). The HMM is a simple synthetic generative process displaying the idea of the data manifold hypothesis ~\citep{bengio_2013}, where data lie on $D$-dimensional submanifold of the ambient $N$-dimensional space. This generative process has been introduced and investigated in ~\cite{goldt2020modeling,goldt2022gaussian,gerace2020generalisation}. We stress that there are other works in the literature that employ similar data models, such as~\cite{boffi}, while studies contained in ~\cite{ross, chen_score_2023, pidstrigach_score-based_nodate, stanczuk_your_2023} study the effect of a latent data dimensionality on generative diffusion.
According to the HMM, data points $\{\xi^{\mu}\in \mathbb{R}^N\}_{\mu = 1}^P$ are generated as $\xi^{\mu}=g\left(\frac{1}{\sqrt{D}}F z^{\mu}\right)$, where the latent variables $z^\mu$ are Gaussian, $z^{\mu}\sim\mathcal{N}(0,I_{D})$, $g$ is an element-wise
non-linearity, and $F\in\mathbb{R}^{N\times D}$ is a random matrix with i.i.d. standard Gaussian entries. The number of data points is $P = e^{\alpha N}$, with $\alpha$ a control parameter of the model. We define
$\alpha_D = D/N$, and assume $D,N\to+\infty$ with $\alpha_D$ staying finite. 

\section{Diffusion Models}
\label{sec:diffusion}
Diffusion Models (DMs) are state-of-the-art generative models. These models are capable of generating new examples (e.g. images, videos) through a stochastic dynamical denoising process, occurring in time. Previous works in literature show that data features are progressively learned by DMs during a noising process, which is then reflected in the way they sample, during the de-noising procedure. The REM formalism is a powerful tool to explain such phenomenology, as showed by \cite{biroli_dynamical_2024}.
After introducing the physics of DMs, we are going to tackle the context of highly structured data, specifically focusing on two - apparently complementary - aspects of the model performance:
\begin{itemize}
    \item \textbf{Memorization}: the predisposition of the model to collapse onto the training-data in the last stage of the denoising process. We study how the tendency of these models to memorize change when data live on a latent manifold of a given dimension.
    \item \textbf{Generalization}: the capability of the model to learn the ground-truth distribution of the training-data. We implement the same techniques employed to study memorization to compute the optimal amount of denoising that is necessary to fit the data.
\end{itemize}

\subsection{Forward and Reverse}

We review the generative denoising diffusion formalism as formulated in Ref. \cite{song2021score} in terms of SDEs. We call $p_0$ the target distribution in the $N$-dimensional space, that is, the distribution that we want to generate samples from. $p_0$ is typically unknown, but i.i.d. samples from it are available.

The \emph{forward process} is defined as an SDE with initial condition $x_{t=0}\sim p_{0}$ that evolve according to
\begin{equation}
dx_t= dW_{t},
\label{eq:forward}
\end{equation}
where $W_{t}$ is a Wiener process and the process is integrated up to some final time $t_f\gg1$. We call $p_t(x_t)$ the marginal distribution density of $x_t$ at time $t$. This prescription for the forward process is known as variance exploding, in contrast to the variance-preserving one that leads to a standard Gaussian at large times, while here approximately $x_{t_f}\sim\mathcal{N}\left(0,t_f I_N\right)$.

The \emph{reverse process} instead, goes back in time starting from $x_{t_f}\sim\mathcal{N}\left(0,t_f I_N\right)$,
and according to
\begin{equation}
dx_t=-\nabla_{x}\log p_{t}(x)dt+ dW_{t}
\end{equation}
which takes time reverse from $t_{f}$ to $t=0$. The drift
term $S(x,t)=\nabla_{x}\log p_{t}(x)$ is called score function. It turns out that the forward and the reverse process have the same marginal density $p_t(x_t)$ \cite{song2021score}. In particular, starting from pure noise, the reverse process can be integrated down to time 0 to produce new samples from $p_0$.

We will now consider the case in which $p_0$ is given by Hidden Manifold Model.

\subsection{The True Score Function}
\label{sec:true}

Usually, the true data distribution is not known. In our synthetic setting, though, it can be explicitly written as 
\begin{equation}\label{eq:true-density}
   p_{0}(x)=\int Dz\ \delta\left(x-g\left(\frac{Fz}{\sqrt{D}}\right)\right)\ ,
\end{equation}
where $\int Dz$ is integration with standard Gaussian density in $D$ dimensions.
Therefore, the density of the process at a given time $t$ takes the form
\begin{equation}
    p_{t}(x)=\int Dz\ \frac{1}{\sqrt{2\pi t}^{N}}\ e^{-\frac{1}{2 t}\|x-g\left(\frac{Fz}{\sqrt{D}}\right)\|^{2}}.  
\end{equation}
The score function can be obtained exactly from this expression in the case of linear activation, as shown in \cite{ventura2024spectral}.

\subsection{The Empirical Score Function}
\label{sec:emp}

If we consider the empirical score function, the starting measure is $p_{0,\mathcal{D}}^{emp}(x)=\frac{1}{P}\sum_{\mu=1}^{P}\delta\left(x-\xi^{\mu}\right)$.
After time $t$, the forward process generates points distributed according to the probability $p_{t}(x)$, whose empirical approximation is 
\begin{equation}
\label{eq:emp_pdf}
p_{t,\mathcal{D}}^{emp}(x)=\frac{1}{P\sqrt{2\pi t}^{N}}\sum_{\mu=1}^P e^{-\frac{1}{2 t}\|x-\xi^{\mu}\|^{2}}.
\end{equation}

\section{The Random Energy Model formalism}
\label{sec:REM}
In order to compute the main quantities that characterize Diffusion Models (DMs), we introduce the tools needed to solve a generic REM, following \cite{lucibello_exponential_2023}. 

Let us consider $P=e^{\alpha N}$ (or equivalently $P=e^{\alpha N}-1$) i.i.d. energy levels $\ve^\mu \sim p(\ve \,|\, \omega)$, where we extend the typical REM setting allowing for a common source of quenched disorder $\omega \sim p_\omega$. 
The goal is to compute the average asymptotic free energy of the system, defined by
\begin{equation}
\label{eq:free_ene}
\phi_\alpha(\lambda) = \lim_{N\to\infty}  \frac{1}{\lambda N} \mathbb{E}\log \sum_\mu e^{\lambda N\ve^\mu}     
\end{equation}

We shall assume that the probability distribution of the energy levels is such that, with probability one over the choice of $\omega$ when $N\to \infty$ the cumulant generating function has a well defined limit:
$ lim_{N\to\infty}\frac{1}{N}\log\mathbb{E}_{\ve|\omega}\, e^{\lambda N \ve}$ exists, and the distribution over the choices of $\omega$ concentrates around its mean. Then we define the typical cumulant generating function and its Legendre transform::
\begin{align}
\zeta(\lambda)&=\lim_{N\to\infty}\frac{1}{N}\mathbb{E}_\omega\log\mathbb{E}_{\ve|\omega}\, e^{\lambda N \ve},\label{eq:zeta}\\
s(\ve)&=\sup_{\lambda}\ \ve\lambda-\zeta(\lambda).
\label{eq:s}
\end{align}

The total entropy of the system is $\Sigma(\epsilon)=\alpha-s(\ve)$.
Depending on the value of $\Sigma(\epsilon)$, the REM displays a separation into two thermodynamic phases: an \textit{uncondensed} phase where the system can \textit{populate} an exponential number of energy levels, at lower values of $\lambda$; a \textit{condensed} phase where the system is able to populate a unique energy state, at higher values of $\lambda$.\\

Let us define the quantities $\ve_{*}(\alpha)$ and $\lambda_{*}(\alpha)$
respectively as the 
maximum value of the energy levels 
in the uncondensed phase, obtained as the
largest root of $\Sigma(\ve_{*})=0$, and the condensation
threshold. Notice that we are seeking for the maximum energy, by definition of the free-energy function in Eq. \eqref{eq:free_ene}. In the uncondensed phase, i.e. when $\lambda<\lambda_{*}(\alpha)$,
the dominating energy level $\tilde{\ve}(\lambda)$ is obtained
as the stationary point of $\lambda\ve-s(\ve)$,
and by the Legendre transform definition of $\zeta(\lambda)$
this is equivalent to $\tilde{\ve}(\lambda)=\zeta'(\lambda)$.
The entropy of the dominating state can be rewritten as $\Sigma(\tilde{\ve}(\lambda))=\alpha-s(\tilde{\ve}(\lambda))=\alpha+\zeta(\lambda)-\lambda\zeta'(\lambda)$,
so the condensation threshold $\lambda_{*}(\alpha)$ is obtained from
the condensation condition
\begin{equation}
  \alpha+\zeta(\lambda_{*})-\lambda_{*}\zeta'(\lambda_{*})=0.  
\end{equation}
Finally, the free energy is given by
\begin{equation}
\phi_{\alpha}(\lambda)=\begin{cases}
\frac{\alpha+\zeta(\lambda)}{\lambda} & \lambda<\lambda_{*}(\alpha),\\
\ve_{*}(\alpha) & \lambda\geq\lambda_{*}(\alpha).
\end{cases}\label{eq:phi-rem}
\end{equation}

\section{Memorization in Generative Diffusion}
\label{sec:memo}

\begin{figure}
    \centering
    \includegraphics[width=0.85\linewidth]{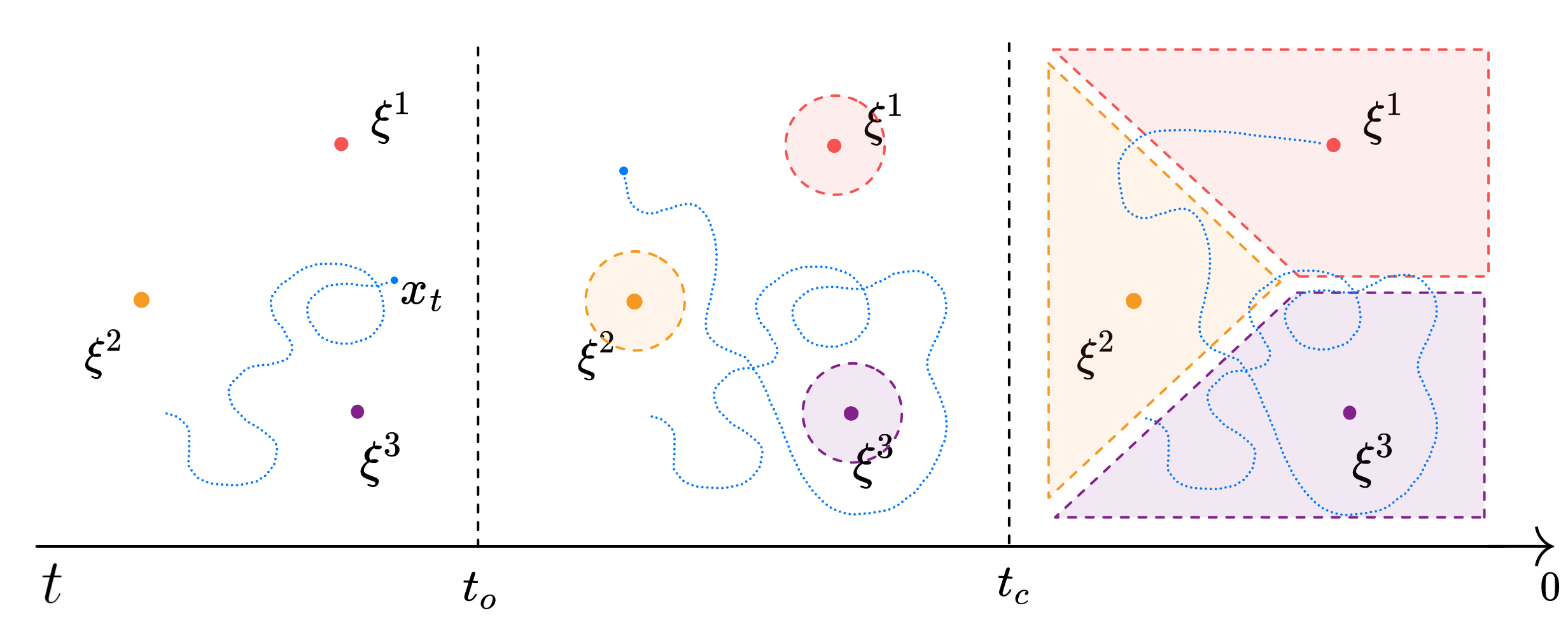}
    \caption{Pictorial representation of the phases identified in the reverse process (from large times to $t=0$) driven by the empirical score. The evolution of a typical trajectory is represented by a dotted blue line. For $t < t_{o}$, data points form a basin around them, but the typical trajectory is not affected by this. For $t<t_{c}$, the basins of attraction of the data points cover the whole space, trajectories cannot escape them once inside, and eventually fall into the data points at time $t=0$.}
    \label{fig:phases}
\end{figure}

We here analyze the memorization phenomenology in generative diffusion when the model is trained on structured data. We will hereby use three expressions that all refer to the same dynamic process: \textit{collapse}, \textit{condensation} and \textit{memorization}. 
The first two idioms, which derive from the REM terminology, will be proved to coincide in this framework, due to the typicality of the stochastic trajectories involved (see \cite{achilli2024losing} for a case where this equivalence does not hold); the third concept, i.e. memorization, is more widely employed in the literature and we will use it as an umbrella term for the first two. Following \cite{biroli_dynamical_2024}, we are treating the attraction of the diffusive trajectories by the data points in terms of the collapse phase-transition occurring in an effective REM. 
We find two main dynamical events occurring in time:
\begin{enumerate}
    \item The appearance of attractors with finite basins of attraction in the diffusion at time $t = t_o$. We call this time \textit{onset time}, and it consists in the moment when training data become attractive, yet without influencing the typical diffusive trajectory of the model. This is also the time where data typically become local maxima of the mixture of gaussian in Eq. \eqref{eq:emp_pdf}.  
    \item The \textit{collapse} of the typical diffusive trajectory on the training data points, occurring at time $t = t_c < t_o$. 
\end{enumerate}
Fig. \ref{fig:phases} provides for a sketch of the phase separation described above.  

\subsection{Collapse Time}

Here we first recap the collapse condition for diffusion models as it was introduced in Ref. \cite{biroli_dynamical_2024}, and then proceed to compute it for our data-generating model. If we start the forward diffusion process from one of the data points, e.g. $\xi^{1}$, then the typical trajectory is  $x_{t}=\xi^{1}+\omega\sqrt{t}$,
with $\omega\sim\mathcal{N}(0,I_N)$. We want to see at which time $t_{c}$
the term $\mu=1$ dominates the summation in the measure, which for
our choice of $x_{t}$ takes the form
\begin{align}
p^{emp}_{t}(x) & =\frac{1}{P\sqrt{2\pi t}^{N}}\left(e^{-\frac{\|\omega\|^{2}}{2}}+\sum_{\mu\geq2}e^{-\frac{1}{2 t}\|(\xi^{1}-\xi^{\mu})+\omega\sqrt{t}\|^{2}}\right)\\
 & =\frac{1}{P\sqrt{2\pi t}^N}\left(Z_{1}+Z_{2,\ldots,P}\right).
\end{align}
In the limit of $P,N\to\infty$ with $\alpha=\frac{\log P}{N}$ fixed,
we find $Z_{1}\simeq e^{-N/2}$, while $ \frac{1}{N}\log Z_{2,...,P}$  concentrates around $\phi_t$, with
\begin{equation}\label{eq:phi-diffusion}
\phi_t = \lim_{N\to\infty}\frac{1}{N}\log\sum_{\mu\geq2}e^{-\frac{1}{2 t}\|(\xi^{1}-\xi^{\mu})+\omega\sqrt{t}\|^{2}}.
\end{equation}
One can make a signal-to-noise argument by comparing the concentrated versions of $Z_1$ and $Z_{2,..,P}$. This approach leads to the so-called \textit{collapse} criterion, also used in Refs. \cite{biroli_dynamical_2024, lucibello_exponential_2023}. This criterion requires
\begin{equation}
\label{eq:collapse}
  \alpha+\zeta_{t_c}(1)=-\frac{1}{2}.  
\end{equation}
Since now the noise in the process is played by the factor $\lambda/t$. As noticeable from Eq. \eqref{eq:collapse}, in this problem we are imposing $\lambda^* = 1$ to compute the time $t_c$ at which collapse occurs. 
For given $\xi^{1}$, $\phi_t$ is minus the average free energy density of
a REM, 
$\phi_t =\lim_{N \to \infty}\frac{1}{N} \log\sum_{\mu\geq2}e^{\epsilon_{\mu}}$
with
$P-1$ energy levels $\epsilon_{\mu}=-\frac{1}{2 t}\|(\xi^{1}-\xi^{\mu})+\omega\sqrt{t}\|^{2}$.

We then need to find the cumulant generating function for the energy levels
\begin{align}
\zeta_t(\lambda) & =\lim_{N\to+\infty}\frac{1}{N}\log\mathbb{E}_{\epsilon}e^{\lambda\epsilon}\\
 & =\lim_{N\to+\infty}\frac{1}{N}\mathbb{E}_{\xi^{1},\omega}\log\mathbb{E}_{\xi^{2}}e^{-\frac{\lambda}{2 t}\|(\xi^{1}-\xi^{\mu})+\omega\sqrt{t}\|^{2}}.
 \label{eq:zeta-diffusion}
\end{align}
If we assume that the data points come from a linear manifold, $\xi^{\mu}=\frac{1}{\sqrt{D}}Fz^{\mu}$, then Eq. \eqref{eq:zeta-diffusion} becomes

\begin{equation}
\zeta_{t}(\lambda)=\lim_{N\to\infty}\frac{1}{N}\mathbb{E}_{F,z^{1},\omega}\log\mathbb{E}_{z^{2}}e^{-\frac{\lambda}{2 t}\lVert\left(Fz^{2}-Fz^{1}\right)+\omega\sqrt{t}\rVert^{2}}.
\end{equation}
In order to investigate the scaling of $t_c$ with respect to the control parameters, let us simplify even more the data model and assume that
$D$ dimensions have variance $\sigma_{i}^{2}=\sigma^{2}$ and $N-D$
have variance $\sigma_{i}^{2}=0$. We have
\begin{equation}
\label{eq:zetaiso}
\zeta_{t}(\lambda)=-\frac{1}{2}\alpha_{D}\log(1+\frac{\lambda}{t}\sigma^{2})-\frac{\lambda}{2}\alpha_{D}\frac{t+\sigma^{2}}{t+\lambda\sigma^{2}}-\frac{\lambda}{2}(1-\alpha_{D}).
\end{equation}
We can find the collapse time from the condition in Eq.~\eqref{eq:collapse}
whose solution is 
\begin{equation}
\label{eq:tc_homo}
t_{c} =\frac{\sigma^{2}N/D}{e^{2\log P/D}-1}.
\end{equation}
Appendix \ref{app:vp} contains the same derivation in the variance-preserving scenario, for comparison with estimates obtained by the unstructured case in \cite{biroli_dynamical_2024}: the main difference, which is conserved in the variance-exploding model, lies in the substitution of the visible dimension $N$ with the latent one $D$ in the exponent contained in $t_c$.
The collapse time depends on the manifold dimension and the number of hidden points. The so-called "curse of dimensionality", i.e. the need for a number of training data points that scales exponentially in the visible dimension of the data-space \cite{yarotzky, cybenko}, has been mitigated by the fact that we have an effective dimensionality for the data.

If we consider the limit of $D\ll\log P$ and $D\ll N$ we have
\begin{equation}
\label{eq:tcasymp}
t_{c}\approx\frac{\sigma^{2}}{2}\frac{N}{D}e^{-\frac{2\log P}{D}},
\end{equation}
which goes to zero fast.

\begin{figure}[t]
    \centering
    \includegraphics[width=0.45\textwidth]{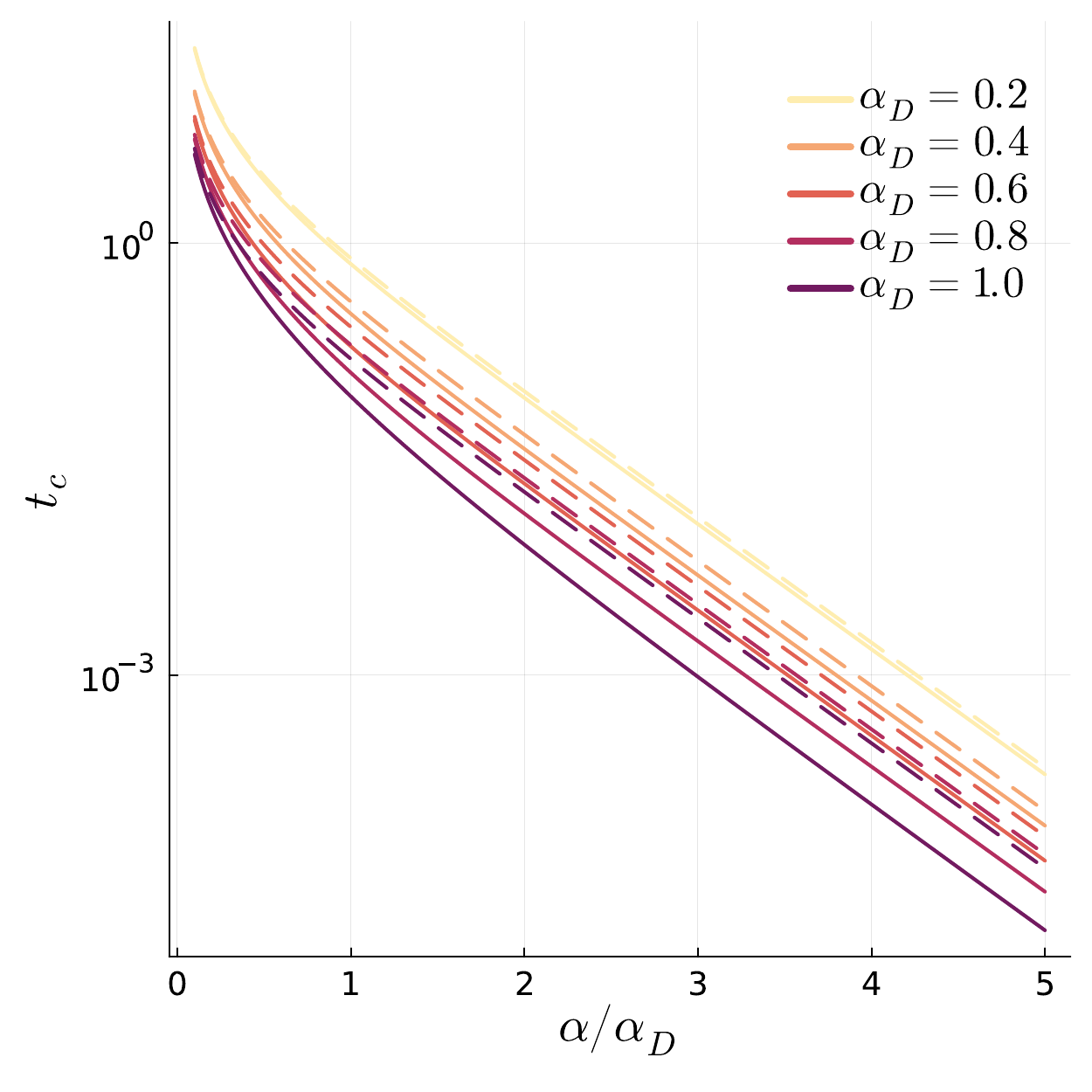}
    \includegraphics[width=0.45\textwidth]{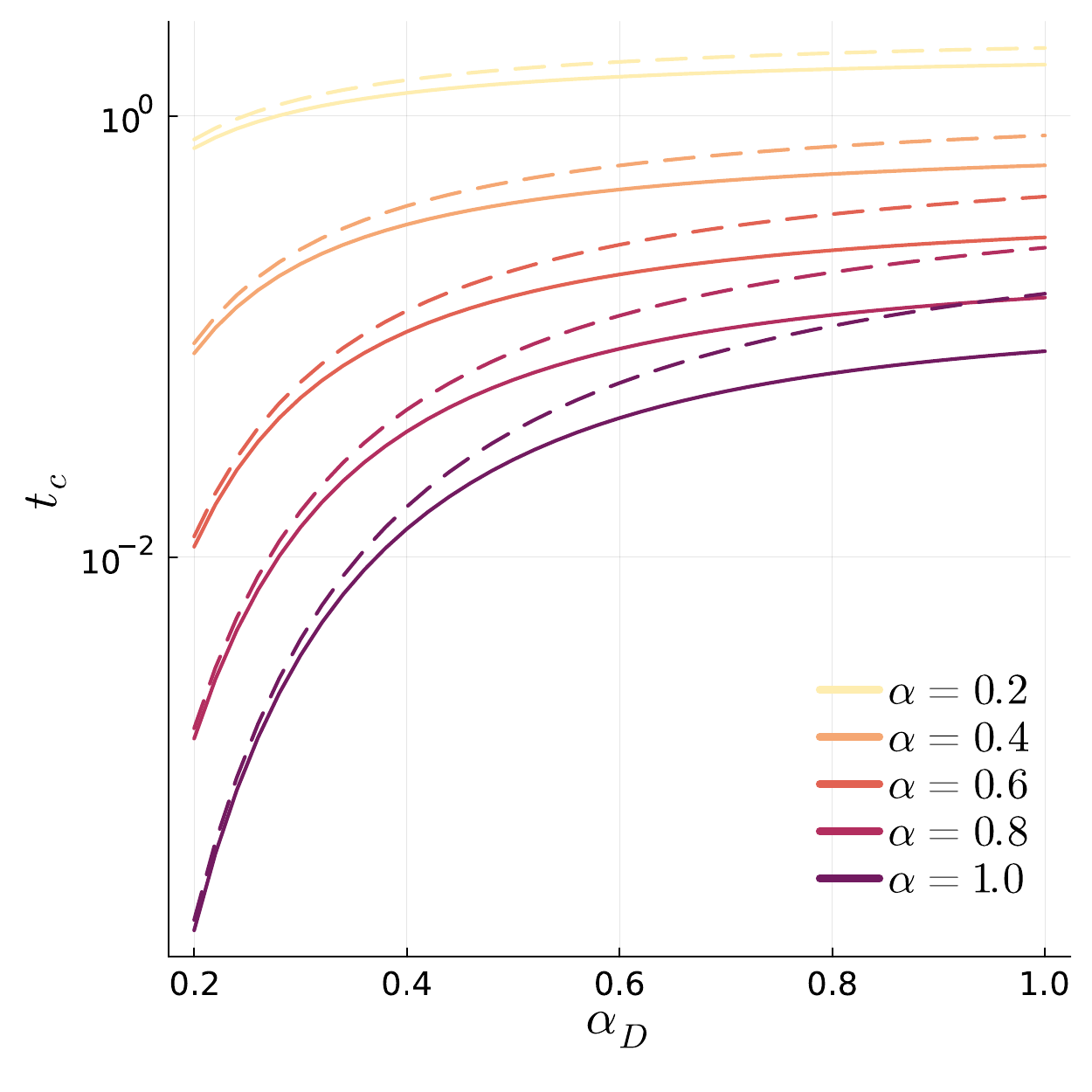}
     \caption{Semi-logarithmic plots of $t_c$ in the linear manifold case (solid) compared to the homogeneous Gaussian case (dashed) for different values of $\alpha_D$ (\textbf{Left}) and $\alpha$ (\textbf{Right}). }
     \label{fig:tc-scaling}
\end{figure}


In the case where the data points come from a linear manifold, we can solve numerically the collapse equation derived in Appendix~\ref{appendix:zeta-linear}. In Fig.~\ref{fig:tc-scaling} we show how $t_c$ scales with the ratio $\alpha/\alpha_D$, i.e. $\log P / D$. These curves are compared with the Gaussian expression for $t_c$ contained in Eq.~\eqref{eq:tc_homo}. It is straightforward to notice that the slopes of the curves are the same for $\alpha \gg \alpha_D$, meaning that even in the linear manifold case we observe the same exponential scaling with $\log P / D$ obtained for the homogeneous Gaussian scenario. 
Fixing $\alpha$, which here corresponds to fixing the number of data points, we see that the collapse time decreases with the hidden dimensionality $D$. Moreover, collapse occurs earlier in the reversed process when the number of data points is smaller. 

\begin{figure}[t]
    \centering
    \includegraphics[width=0.45\textwidth]{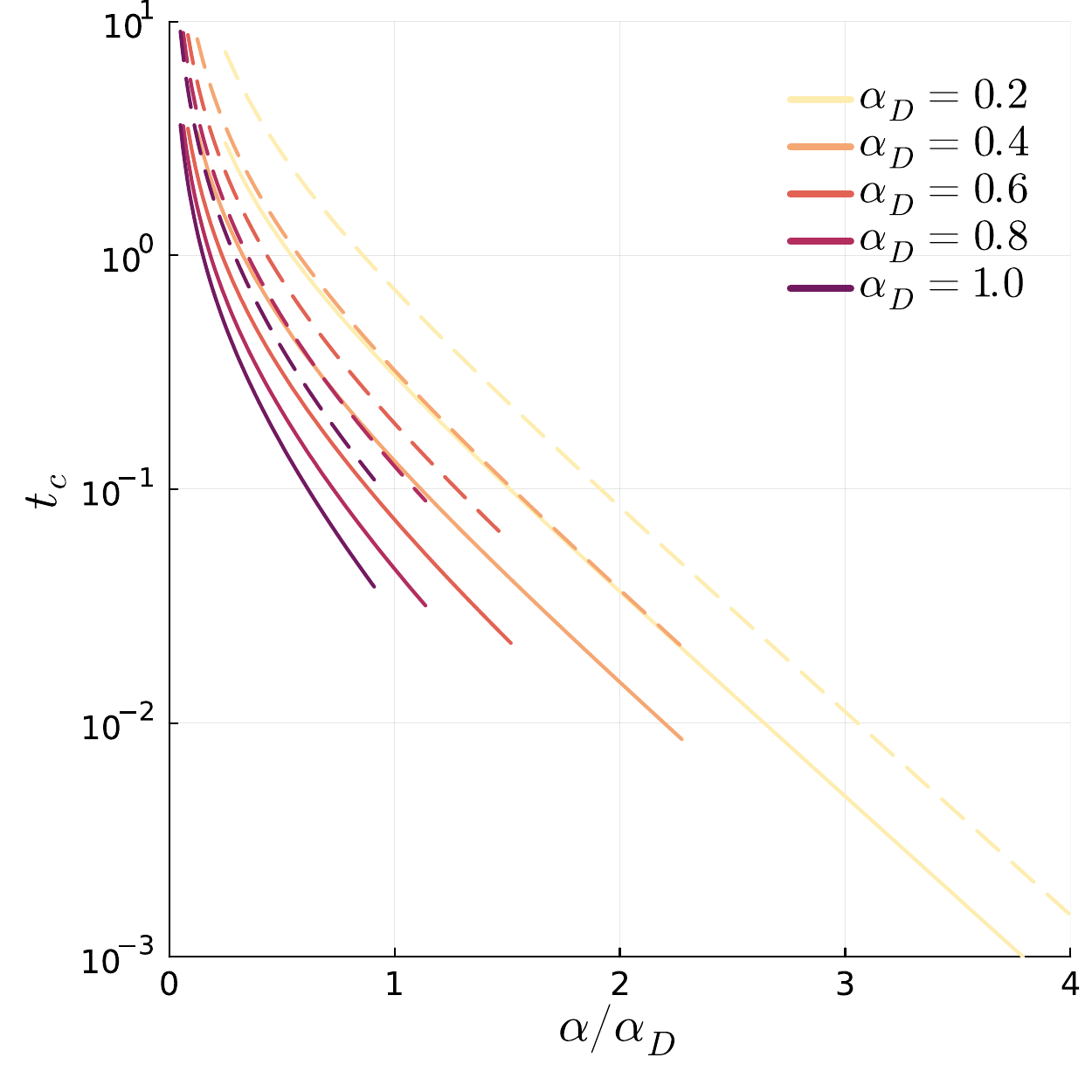}
    \includegraphics[width=0.45\textwidth]{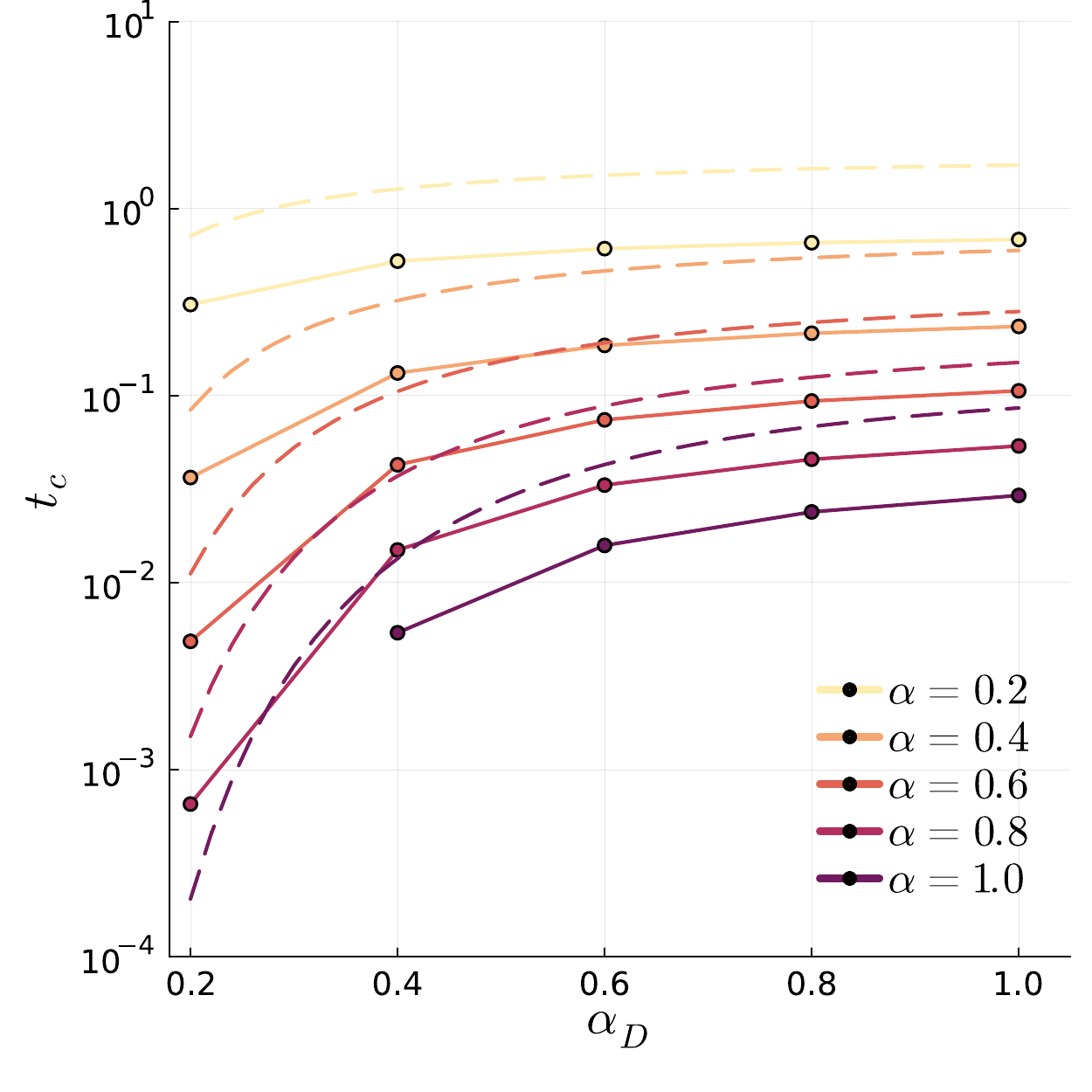}
     \caption{Semi-logarithmic plots of $t_c$ in the hidden manifold  case (solid) with tanh activation compared to the linear manifold case (dashed) for different values of $\alpha_D$ (\textbf{Left}) and $\alpha$ (\textbf{Right}). }
     \label{fig:tc_nl-scaling}
\end{figure}

Let us now consider a non-linear manifold for the data points, i.e. $\xi^{\mu}=g\left(\frac{1}{\sqrt{D}}Fz^{\mu}\right)$. In this case
Eq.~\eqref{eq:zeta-diffusion} assumes the following expression

\begin{equation}
   \zeta_{t}(\lambda)=\lim_{N\to+\infty}\frac{1}{N}\mathbb{E}_{z^{1},F,\omega} \log\mathbb{E}_{z^{2}}e^{-\frac{\lambda}{2 t}\lVert\left(g\left(\frac{1}{\sqrt{D}}Fz^{1}\right)-g\left(\frac{1}{\sqrt{D}}Fz^{2}\right)\right)+\omega\sqrt{t}\rVert^{2}}. 
\end{equation}
This function can be computed using the replica method, as shown in Appendix~\ref{appendix:zeta-non-lin-1}. We find an expression for $\zeta_t$ in the RS approximation, i.e.

\begin{equation}
\zeta_t(\lambda; q_{d},q_{0},m,\hat{q}_{d},\hat{q}_{0},\hat{m})=-\alpha_{D}m\hat{m}-\frac{\alpha_{D}}{2}(q_{d}\hat{q}_{d}-q_{0}\hat{q}_{0})+\alpha_{D}G_{S}(\hat{q}_{d},\hat{q}_{0},\hat{m})+G_{E}(\lambda, t; q_{d},q_{0},m),
\end{equation}
with
\begin{align}
G_{S}(\hat{q}_{d},\hat{q}_{0},\hat{m}) & =-\frac{1}{2}\log\left(1-\hat{q}_{d}+\hat{q}_{0}\right)+\frac{1}{2}\frac{\hat{m}^{2}+\hat{q}_{0}}{1-\hat{q}_{d}+\hat{q}_{0}},
\end{align}
and
\begin{equation}
G_{E}(\lambda, t; q_{d},q_{0},m) =\int D\omega\int D\gamma\int Du^{0}\log\left(\int Du\ e^{-\frac{\lambda}{2t}\left(g\left(u^{0}\right)-g\left(\sqrt{q_{d}-q_{0}}u+mu^{0}-\sqrt{q_{0}-m^{2}}\gamma\right)+\sqrt{t}\omega\right)^{2}}\right).
\end{equation}
Then we solve the saddle point equations (which depend on the choice of the non-linearity) to obtain the typical value of $\zeta_t$. 
At this point, the collapse condition is solved numerically, and the scaling of the collapse time can be compared to the one found for linear manifolds. Fig.~\ref{fig:tc_nl-scaling} depicts the instance of $g(x) = \tanh{(x)}$. As one can notice, curves for the non-linear and linear cases show the same qualitative behavior, displaying the same type of scaling as a function fo $\alpha / \alpha_D$ and $\alpha_D$ when the number of data are fixed.  

\subsubsection{Equivalence between Collapse and Condensation} 
In Eq.~\eqref{eq:collapse} we have introduced a criterion for collapse
time.
In Section~\ref{sec:REM} we have also discussed the condensation threshold for the REM which, in the context of DMs reads
\begin{equation}
\alpha+\zeta_{t_{cond}}(1)-\zeta'_{t_{cond}}(1)=0.
\end{equation}
In order to establish that the condensation and collapse phenomena happen at the same time, $t_{c}=t_{cond}$, we would therefore need
to prove that
\begin{equation}
\zeta'_{t_{c}}(1)=-\frac{1}{2}.
\end{equation}
This is indeed what we find for a typical trajectory as a consequence of the Nishimori condition. Computations are reported in Appendix~\ref{appendix:equivalence}.


\subsection{Onset Time and Basins of Attraction}

The goal of this Section is to compute the onset time $t_o$, i.e. the time at which data points start to become attractors in the diffusion potential. Since the computation does not average over the typical positions sampled by the reverse process, even if data become locally attractive, we find that they do not influence the typical trajectories
until $t_{c}$. 
This aspect is the main difference between the onset time and the \textit{speciation} time computed by \cite{Biroli_2023}: while the former is intrinsic in the dataset itself, the latter depends on the structure of the data points as divided in multiple classes and it does affect the direction of the diffusion in the ambient space.

\begin{figure}
    \centering
    \includegraphics[width=0.32\textwidth]{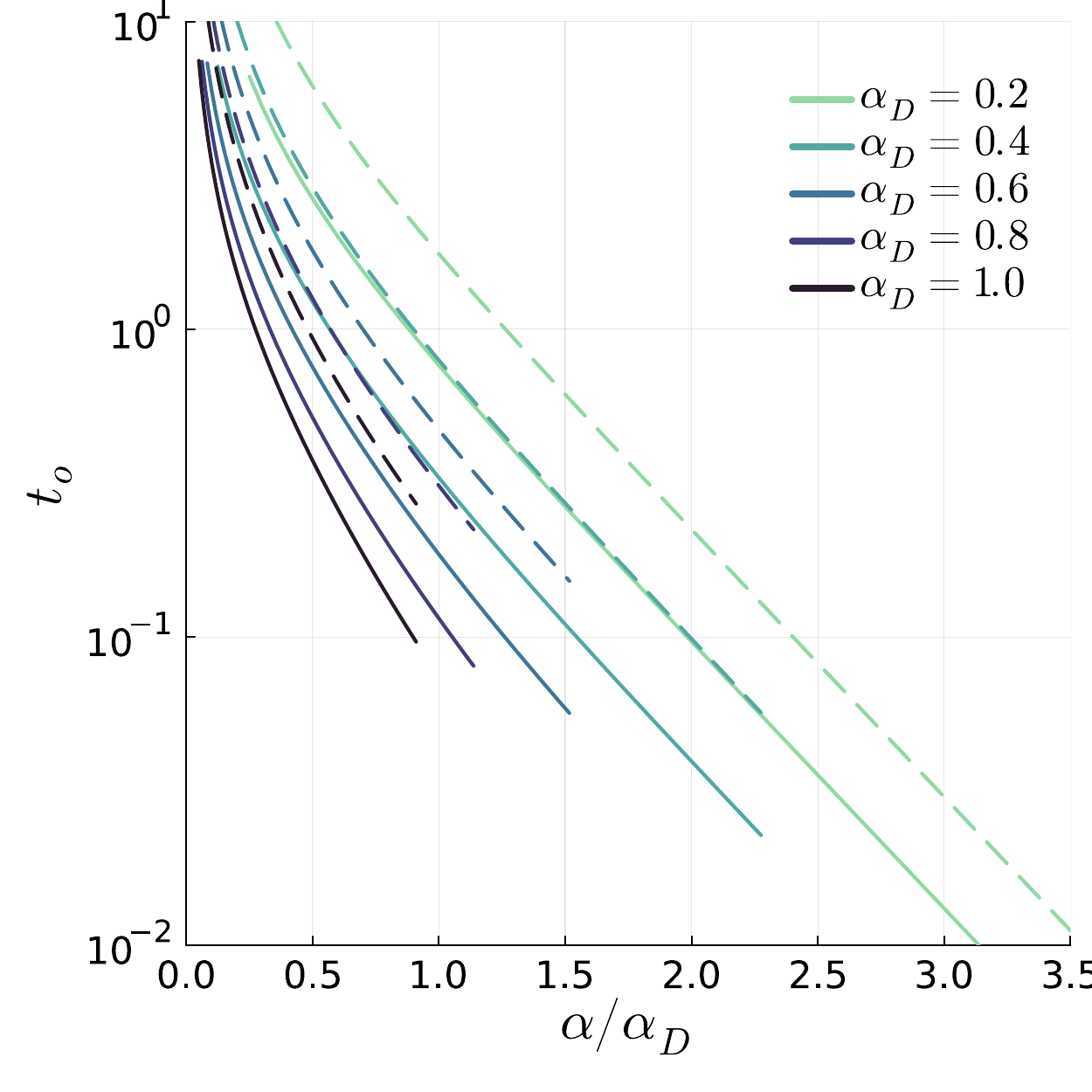}
    \hfill
     \includegraphics[width=0.32\textwidth]{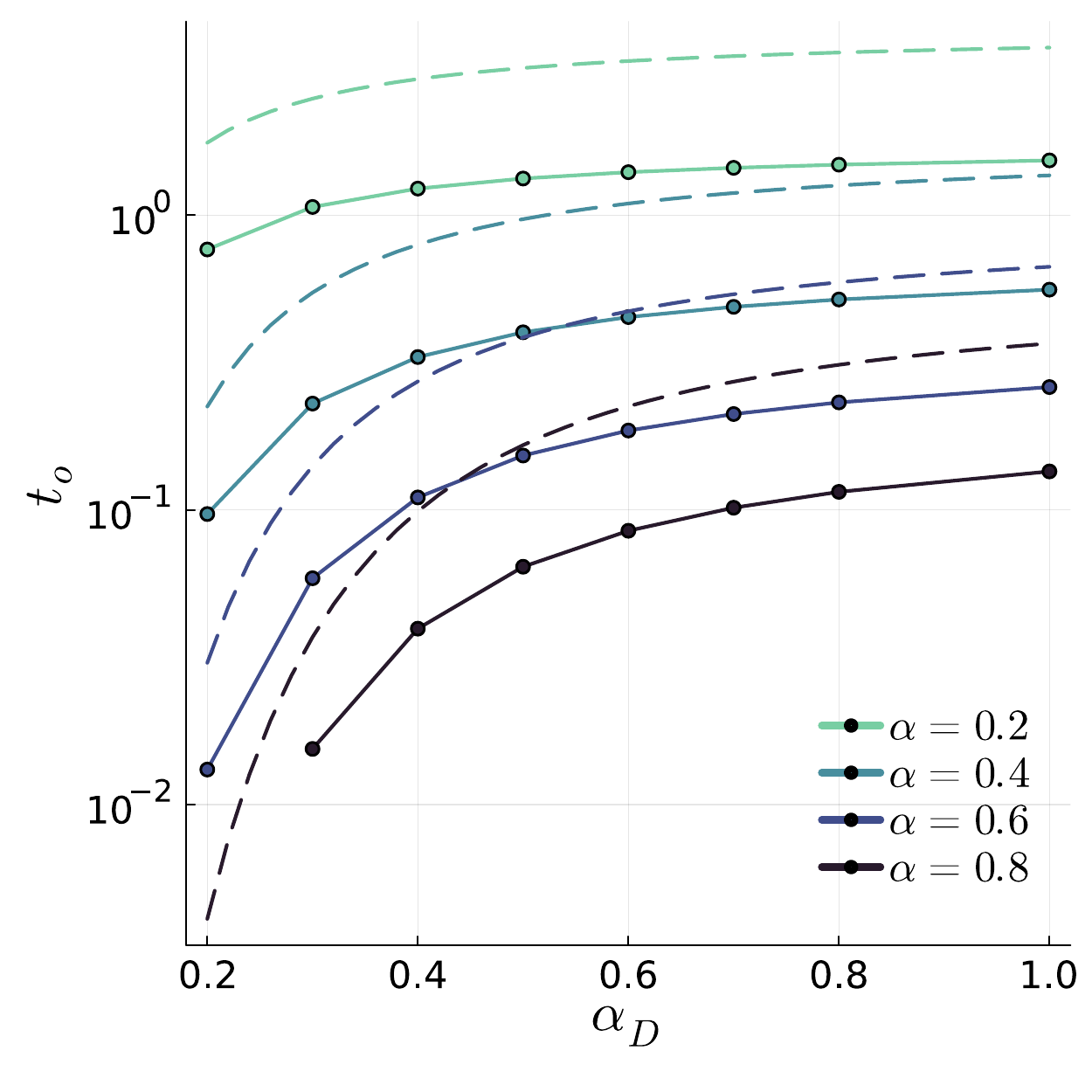}
    \hfill
    \includegraphics[width=0.32\textwidth]{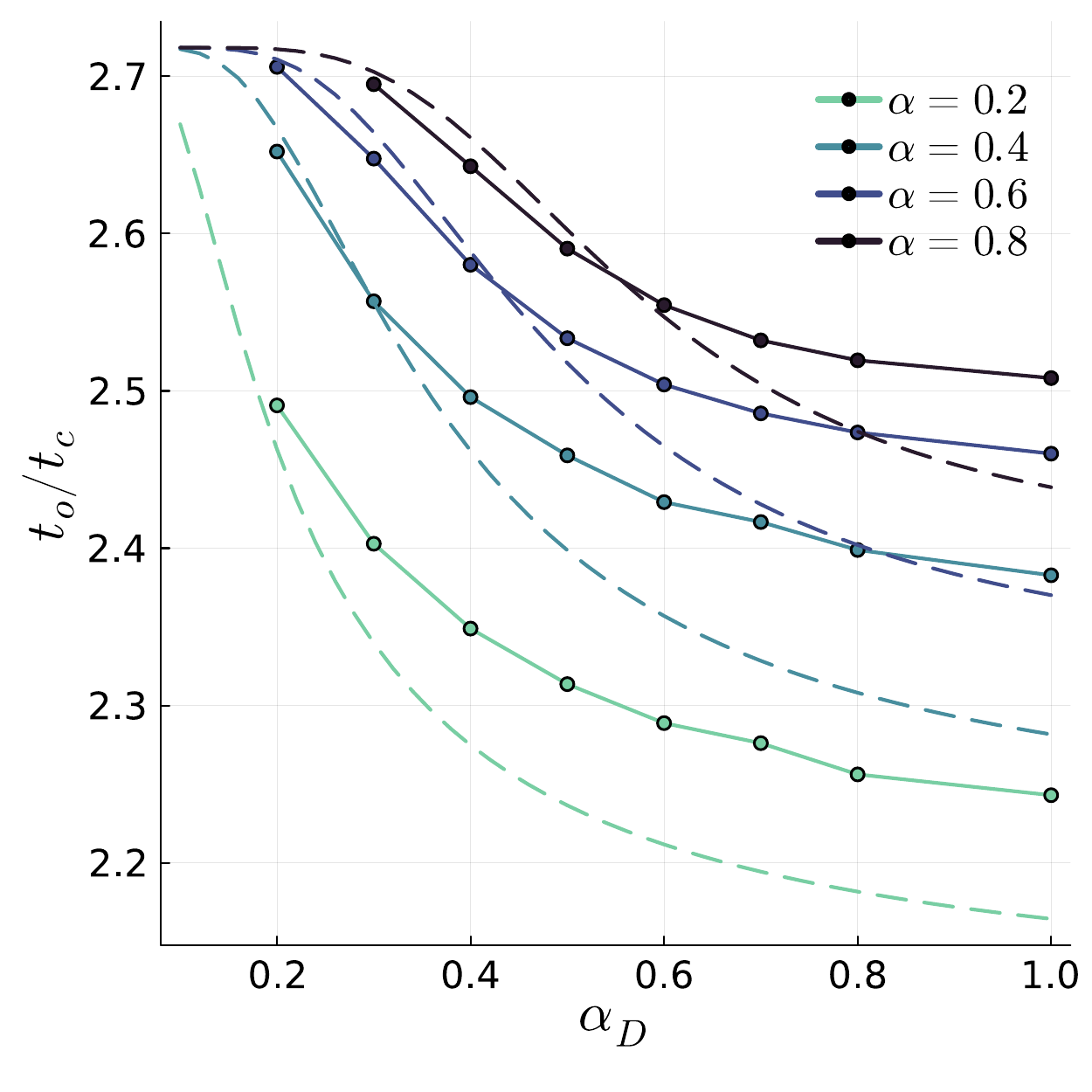}
    \caption{(\textbf{Left}) Onset time $t_o$ as a function of $\alpha / \alpha_D$ in semi-log scale in the hidden manifold  case (solid) with $\tanh$ activation compared to the linear manifold case (dashed); (\textbf{Center}) $t_o$ as a function of $\alpha_D$ for fixed $\alpha$ in semi-log scale for $\tanh$ activation; (\textbf{Right}) comparison of the onset time $t_o$ with the collapse time $t_c$ as a function of $\alpha_D$ when $\alpha$ is fixed and $g = \text{tanh}$.}
    \label{fig:to-nl}
\end{figure}
The onset time can be computed setting $x_{t}=\xi^{1}$
and checking when the corresponding collapse condition is satisfied. Such, condition, that is analogous to the one in Eq.\eqref{eq:collapse}, consists in requiring the relative REM free energy equal to zero, i.e. $\phi_{t_o} = 0$. 
As done for the condensation time, let us first compute the collapse time in the simple homogeneous Gaussian setting, where $D$ variances are equal to $\sigma^{2}$ and the remaining ones are null. The moment-generating function of the relative REM is
\begin{align}
\label{eq:zetaR0}
\zeta_{t}(\lambda) & =\lim_{N\to\infty}\frac{1}{N}\mathbb{E}_{\xi^{1}}\log\mathbb{E}_{\xi}e^{-\frac{\lambda}{2t}\lVert\xi^{1}-\xi\rVert^{2}}
 =-\frac{\alpha_D}{2}\left(\log\left(1+\frac{\lambda \sigma^{2}}{\alpha_{D}t}\right)+\frac{\lambda \sigma^{2}}{\alpha_{D}t + \lambda \sigma^2}\right).
\end{align}
In analogy with the collapse condition in Eq. \eqref{eq:collapse}, the on-set time condition must be
\begin{equation}
\zeta_{t_o}(1)+\alpha=0,
\end{equation}
which reads
\begin{align}
\log\left(1+\frac{\sigma^{2}}{\alpha_{D}t_o}\right)+\frac{\sigma^{2}}{\sigma^2 + \alpha_{D}t_o}	-\frac{2\alpha}{\alpha_{D}}=0.
\end{align}
The same calculation in then performed in the case of manifold structured data for different choices of the $g$ function. The computation is carried out by means of the replica method and it is reported in Appendix \ref{sec:ot}. Results are reported in Fig. \ref{fig:to-nl}. The left panel in the figure shows the onset time as a function of the ratio $\alpha/\alpha_D$, suggesting that $t_o$ behaves similarly to the condensation time $t_c$. The right panel shows how $t_o/t_c$ increases when the data are more structured (i.e. when $\alpha_D$ decreases).
Surprisingly, this quantity also reaches a constant value when $\alpha$ is fixed and $\alpha_D \rightarrow 0$, in the linear case. Due to numerical limitations in computing the saddle point equations we could not observe the same trend in the non-linear scenario. Yet, the good qualitative agreement between the linear and non-linear cases at higher values of $\alpha_D$ suggests that the non-linear model might reach the same plateau. This particular behavior of the onset time might be attributable to the exponentially large size of the basins of attraction of the data points.
\begin{figure}
    \centering
    \includegraphics[width=0.45\linewidth]{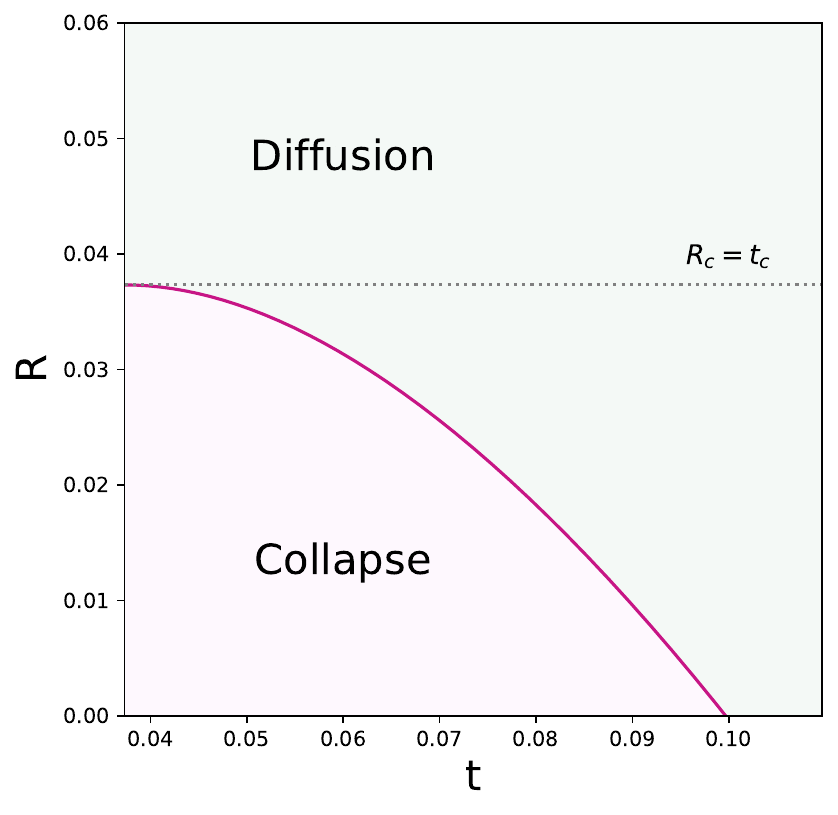}
    \caption{The violet line gives the radius $R$ of the basin of attraction around one data point as a function of time. All particles at a distance smaller than $R(t)$ collapse to the same data point, while the ones at larger distances do not. The basin of attraction appears at $t = t_o$ and it is maximal at $t = t_c$, where the value of $R$ equals the diffusion noise.  Typical trajectories are trapped into the basin of attraction at times smaller than $t_c$. The parameters are $\sigma^2 = 1,\alpha = 1, \alpha_D = 0.5$ .}
    \label{fig:basins}
\end{figure}

Let us now consider a more general case where $x_{t}=\xi^{1}+\omega\sqrt{R}$ where $\omega\sim\mathcal{N}(0,I_N)$ and $R$ is an arbitrary positive real value.
Then one can repeat the calculation for the homogeneous Gaussian framework and obtain
\begin{align}
\label{eq:zetaR}
\zeta_{t,R}(\lambda) & =\lim_{N\to\infty}\frac{1}{N}\mathbb{E}_{\xi^{1},\omega}\log\mathbb{E}_{\xi}e^{-\frac{\lambda }{2t}\lVert(\xi^{1}-\xi)+\omega\sqrt{R}\rVert^{2}}\\
 & =-\frac{1}{2}\left(\alpha_{D}\log\left(1+\frac{\lambda \sigma^{2}}{\alpha_{D}t}\right)+\frac{\alpha_{D}\lambda\sigma^2}{t}\frac{(t - \lambda R)}{\alpha_{D}t+\lambda\sigma^{2}}+\frac{\lambda R}{t}\right).
\end{align}
Note that this expression for $\zeta$ coincides with Eq. \eqref{eq:zetaiso} when $R = t$ and with Eq. \eqref{eq:zetaR0} when $R = 0$.
The new collapse condition for $R(t)$ is given by
\begin{equation}
    \zeta_{t,R_c}(1)+\alpha = -\frac{R_c}{2t}.
\end{equation}
The value of $R_c$ when $t \in [t_c,t_o]$ represents the main distance at which particles would start feeling the attraction to the data point $\xi^1$, i.e. the particle is in the basin of attraction of the pattern if $R<R_c$. Fig. \ref{fig:basins} reports the size of the basins of attraction as a function of the time for one realization of $\sigma^2, \alpha, \alpha_D$. The radius $R$ starts assuming non-zero values at $t = t_o$ and equals the noise of stochastic process $R_c = t_c$ when $t = t_c$. When $t \in [0,t_c]$ each possible trajectory (both typical and non-typical) has collapsed in one of the basins, by definition of collapse time.

\section{Generalization in Generative Diffusion}
\label{sec:generalization}

In this Section we compute the optimal time $t_g$ such that the empirical probability distribution of a DM better fits the target distribution. The degree of generalization of a DM driven by its empirical score can be quantified in terms of the Kullback-Leibler (KL) divergence between the empirical probability distribution of the model and the distribution of the data points on the manifold. 
We first show that the true score and the empirical one do not differ, in the large volume limit, above the collapse transition. Secondly, we calculate $t_g$ for different choices of $\alpha$ and $\alpha_D$, showing that this time is always contained within the condensed phase of the auxiliary REM, i.e. the memorization phase of the DM. A similar effect has been found when seeking the best kernel to approximate probability densities from large-dimensional data: the optimal kernel width is found in the condensed phase \cite{biroli_kernel_2024}.
This is no coincidence: in generative diffusion, the effective probability distribution $ p_t^{emp}$  is a sum of Gaussian kernels centered on the data points.
Finally, since the computation of $t_g$ relies on the presence of collapse over the training set, which is not always encountered in real-world applications of generative diffusion, we propose an alternative criterion to define generalization in DMs.  

\subsection{True vs Empirical Distribution}
\label{sec:true_emp}
\begin{figure}
    \centering
    \includegraphics[width=0.7\linewidth]{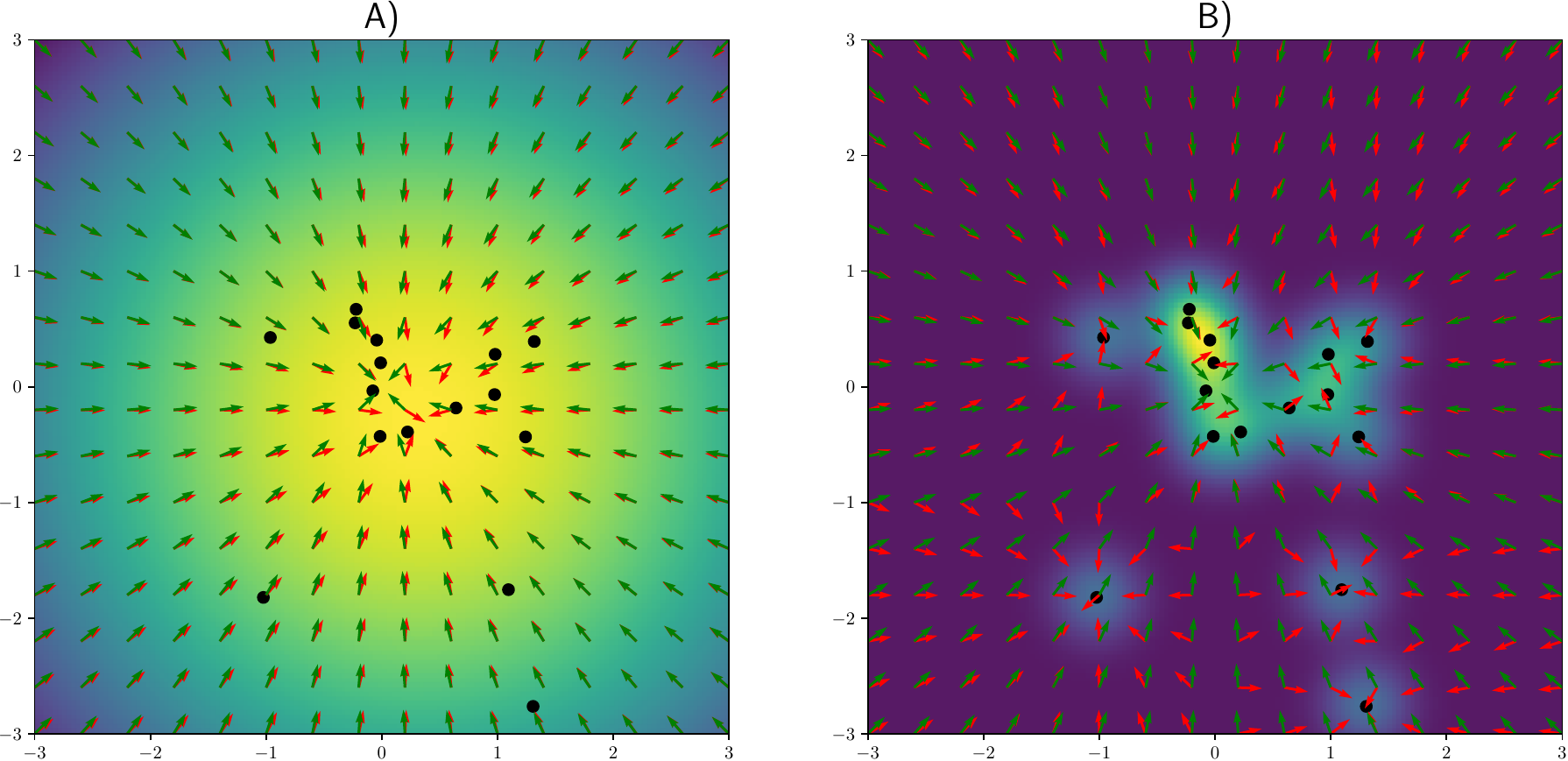}
    \caption{In these Figures, at two given times, the heat map indicates the empirical sampling distribution of a DM, red arrows represent the empirical score while  green ones represent the exact score. The black dots denote individual data points. Panel A depicts such quantities at a $t > t_c$ and panel B displays the typical scenario at a $t < t_c$.  The score transitions from a phase where its direction is dominated by the expectation (i.e. the exact score) to a phase where its orientation is mostly determined by the individual data points.}
    \label{fig:score-discrepancy}
\end{figure}

The Kullback-Leibler (KL) divergence between the true and empirical distribution is 
\begin{equation}
\lim_{N\to\infty}\frac{1}{N}\mathbb{E}_{\mathcal{D}}\,D_{KL}[p_{t}(x)\,|\,p_{t,\mathcal{D}}^{emp}(x)]=\lim_{N\to\infty}\frac{1}{N}\mathbb{E}_{\mathcal{D}}\left[\int dx_{t}\ p_{t}(x)\,\log\,p_{t}(x)-\int dx\ p_{t}(x)\,\log\,p_{t,\mathcal{D}}^{emp}(x)\right]
\end{equation}
In the uncondensed phase we can exploit the fact that the annealed
approximation holds, combined with $\mathbb{E}_{\mathcal{D}}\left[p_{t,\mathcal{D}}^{emp}(x)\right]=p_{t}(x)$
to obtain
\begin{equation}
\label{eq:annealed}
\lim_{N\to\infty}\frac{1}{N}\mathbb{E}_{\mathcal{D}}\,D_{KL}[p_{t}(x)\,|\,p_{t,\mathcal{D}}^{emp}(x)]=\begin{cases}
0 & \text{uncondensed phase}\\
\ve^{*}(t,\alpha)-\alpha-\frac{1}{2}\log(2\pi t)-H_{t} & \text{condensed phase}
\end{cases}
\end{equation}
with $\ve^{*}(t,\alpha)=-\lim_{N\to\infty}\mathbb{E}_{x,\mathcal{D}}\,\frac{1}{2N t}\lVert x_t-\xi^{*}(x_t,\mathcal{D})\rVert^{2}$
and $\xi^{*}$ being the nearest neighbor to $x$ among the data points, while $H_{t}$ is an additional time dependent term. The divergence between the empirical and true scores starting from $t_c$ is represented in the bi-dimensional plot contained in Fig. \ref{fig:score-discrepancy} for one explanatory diffusion experiment, and it is validated by Fig. \ref{fig:kl-tanh} relative to a further analysis of generalization. 

\subsection{Generalization Time: Generalizing while Collapsing}

We would like to understand if there is a time at which the empirical score function points towards the original data manifold and not directly to the data points. 
To study this, we compute the KL divergence between the target distribution, i.e. $p_0$, and the empirical distribution at time $t$, and then minimize it to find the \emph{generalization} time.

\begin{equation}
\lim_{N\to\infty}\frac{1}{N}\mathbb{E}_{\mathcal{D}} D_{KL}[p_{0}\,|\,p_{t,\mathcal{D}}^{emp}] =\lim_{N\to\infty}\frac{1}{N}\mathbb{E}_{\mathcal{D}}\left[ \int dx \ p_{0}(x) \log p_0(x) -\int dx\ p_{0}(x)\,\log\,p_{t,\mathcal{D}}^{emp}(x)\right].
\end{equation}
The second term can be computed using the REM formalism (see Appendix \ref{appendix:kl}) as

\begin{equation}
\label{eq:second_piece}
    \tilde{D}_{KL}[p_0|p_t^{emp}]=-\lim_{N\rightarrow \infty}\frac{1}{N}\mathbb{E}_{\mathcal{D}}\int dx\, p_{0}(x)\,\log\,p_{t,\mathcal{D}}^{emp}(x) \simeq -\phi_{t, \alpha}(1) + \alpha + \frac{1}{2}\log(2\pi t).
\end{equation}

\begin{figure}[ht]
    \centering
    \includegraphics[width=0.6\linewidth]{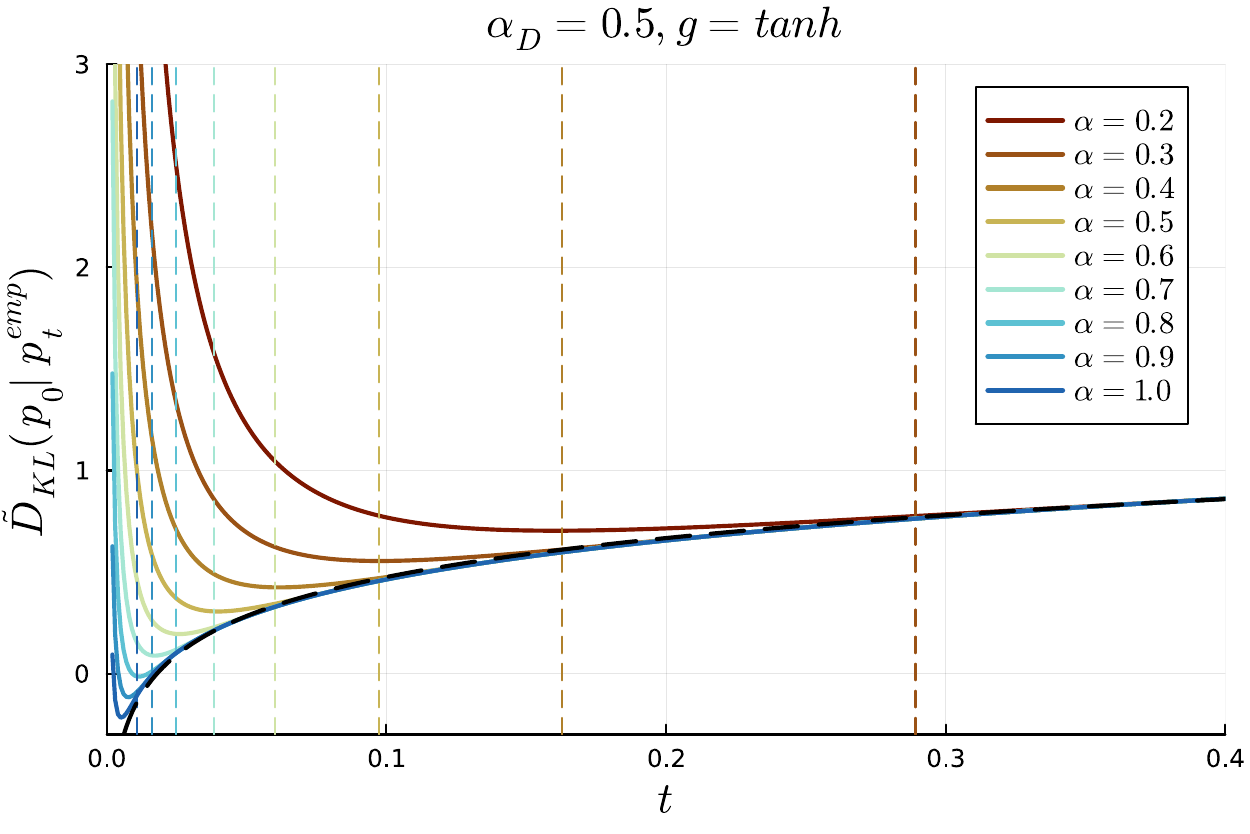}
    \caption{Time-dependent component of the KL divergence between target distribution and empirical distribution as a function of time $t$ and for different values of $\alpha$. The data are generated from a HMM with $\tanh$  activation and aspect ratio $\alpha_D=0.5$. We report with colored dashed lines the condensation time $t_c$ at the corresponding value of $\alpha$, and with the black dashed line the limit $\alpha\to +\infty$.}
    \label{fig:kl-tanh}
\end{figure}

\begin{figure}
    \centering
    \includegraphics[width=0.32\textwidth]{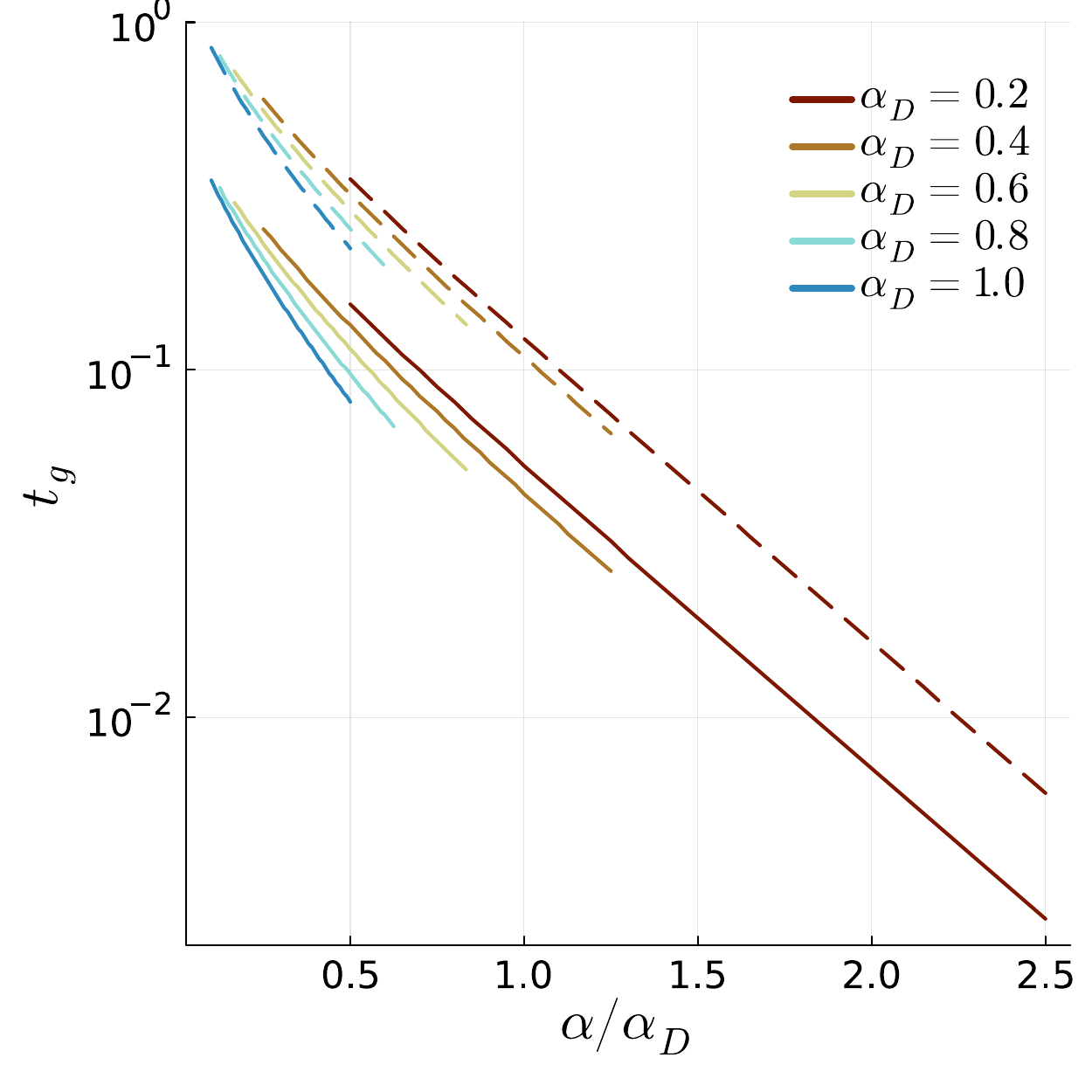}
    \hfill
     \includegraphics[width=0.32\textwidth]{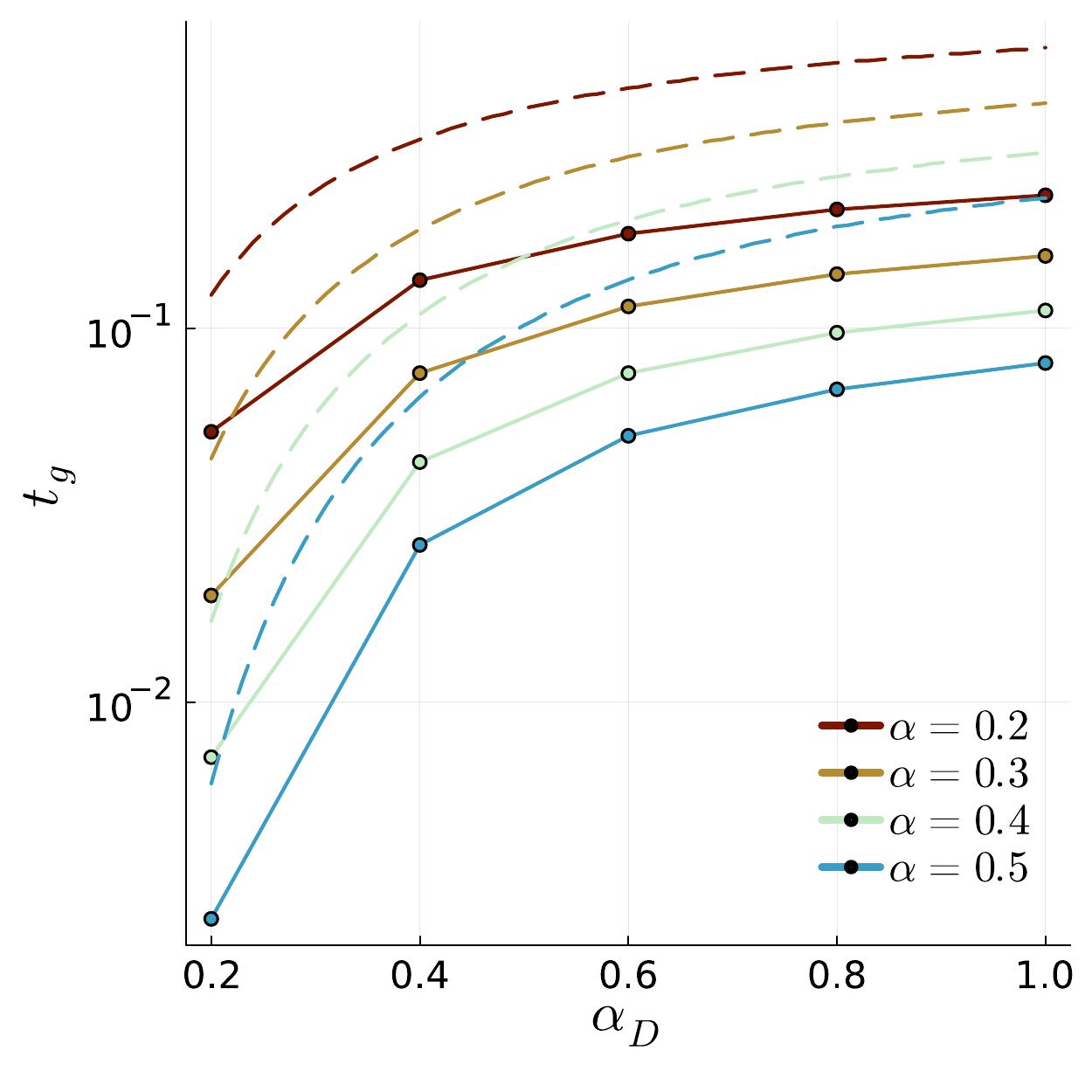}
    \hfill
    \includegraphics[width=0.32\textwidth]{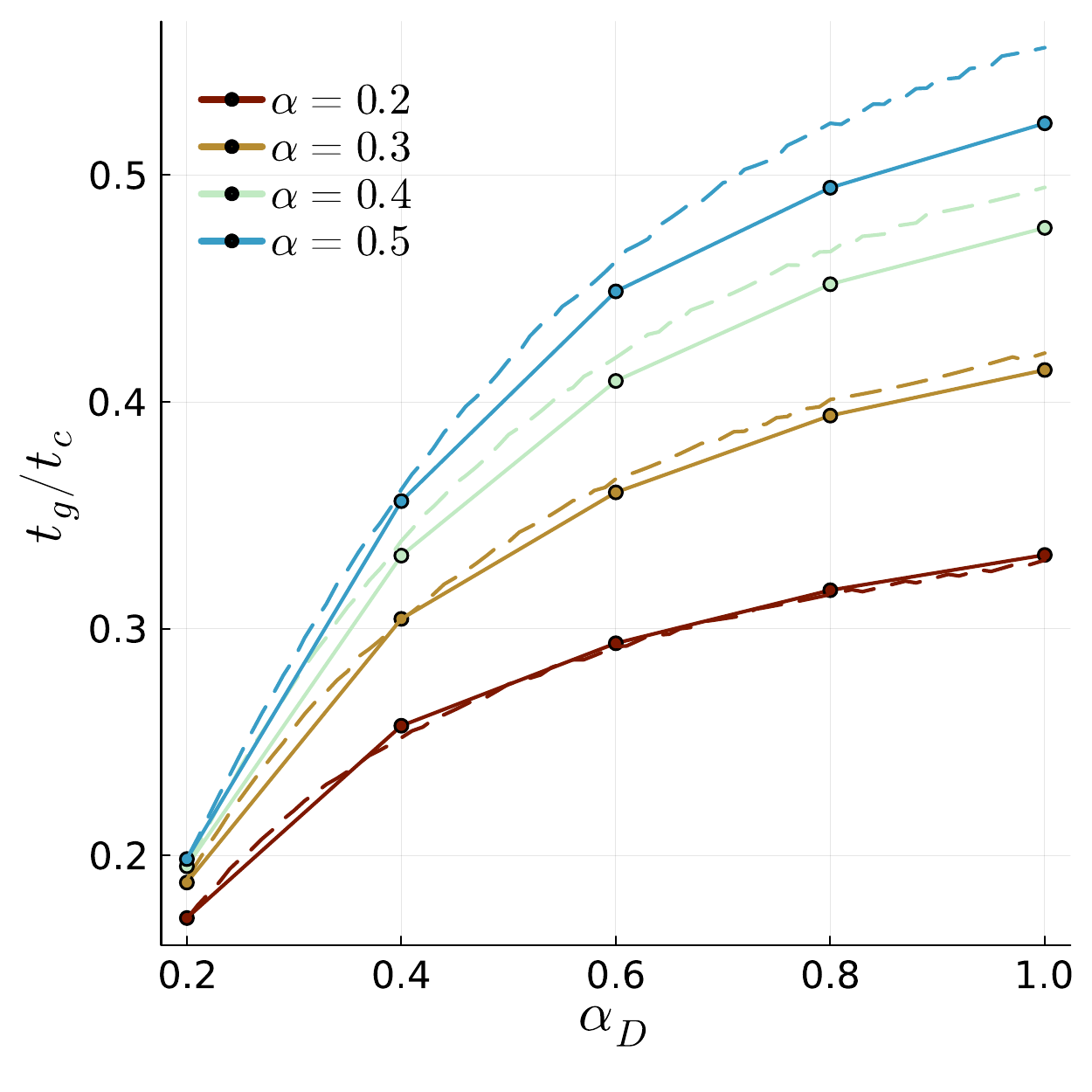}
    \caption{(\textbf{Left}) Generalization time $t_g$ as a function of $\alpha / \alpha_D$ in semi-log scale for $\tanh$ (solid) and linear (dashed) activation; (\textbf{Center}) $t_g$ as a function of $\alpha_D$ for fixed $\alpha$ in semi-log scale for $\tanh$ (solid) and linear (dashed) activation; (\textbf{Right}) comparison of the generalization time $t_g$ with the collapse time $t_c$ as a function of $\alpha_D$ when $\alpha$ is fixed for $\tanh$ (solid) and linear (dashed) activation.}
    \label{fig:tgtc}
\end{figure}

We show the behavior of the KL divergence for data from a hidden manifold model with $\tanh$ non-linearity in Fig.~\ref{fig:kl-tanh}. 
Interestingly, the time $t_g$ where the discrepancy between $p_0$ and $p_{t}^{emp}$ reaches a minimum is always smaller than the corresponding collapse time (reported as a dashed line in the Figure): 
the best generalization of the DM is reached inside the condensation phase, while the diffusive trajectory is trapped into the basin of attraction of the closest data point. 
It is also worth to notice that $$\lim_{\alpha \rightarrow \infty}\tilde{D}_{KL}[p_0|p_t^{emp}] = \tilde{D}_{KL}[p_0|p_t]$$ where $p_t(x)$ is the exact probability distribution of the diffusive process. 
This quantity is represented by the line onto which all the curves in Fig.~\ref{fig:kl-tanh} collapse, i.e. the black dashed line in the figure: the computation in Eq.~\eqref{eq:annealed} is validated by the fact that curves start diverging from the asymptotic line exactly at $t = t_c(\alpha)$.   
Moreover, Fig.~\ref{fig:tgtc} (Center) displays that $t_g$ decreases with $\alpha_D$ when $\alpha$ is fixed, while Fig.~\ref{fig:tgtc} (Right) shows that the ratio $t_g/t_c$ vanishes when $\alpha_D \rightarrow 0$.  This result means that $t_g$ goes to zero faster than the collapse time $t_c$. We can thus conclude that a high structure of the data helps the empirical-score-driven diffusion model for two reasons:
\begin{itemize}
    \item Both $t_c$ and $t_g$ are pushed towards $t = 0$ when $\alpha_D \rightarrow 0$ but the generalization time is moving faster towards smaller times. Since $t_g$ represents the best stopping time to sample along the reverse process
    from the point of view of the KL divergence, one sees that this optimal time occurs after the condensation threshold and much closer to $t = 0$, when the true memorization occurs.
    \item The generalization time occurs inside the memorization phase, i.e. 
    \[
    0 < t_g< t_c \quad \forall \alpha, \alpha_D,
    \]
    and the Kullback-Leibler distance between $p_0$ and $p_t^{emp}$ is a monotonic function in $t\in[t_g,t_c]$ i.e. 
\begin{equation}
    D_{KL}[p_0|p_{t_c}^{emp}]>D_{KL}[p_0|p_{t_g}^{emp}] > 0.
\end{equation}
    Since the empirical model tends to the exact one when $\alpha_D \rightarrow 0$ i.e.
    \begin{equation}
    \lim_{\alpha_D \rightarrow 0}\lim_{N\to\infty}\frac{1}{N}\mathbb{E}_{\mathcal{D}} D_{KL}[p_{0}\,|\,p_{t_c,\mathcal{D}}^{emp}] = 0,   
    \end{equation}
    then we must have 
    \begin{equation}
    \lim_{\alpha_D \rightarrow 0}\lim_{N\to\infty}\frac{1}{N}\mathbb{E}_{\mathcal{D}} D_{KL}[p_{0}\,|\,p_{t_g,\mathcal{D}}^{emp}] = 0,
    \end{equation}
    which means that the degree of generalization of the DM improves when data are more structured.
\end{itemize}

\subsection{Generalization Condition: Generalizing before Collapsing}
We now propose a more empirical definition of generalization for DMs. The main idea consists of sampling configurations from the data-manifold before the model enters its memorization phase. The current definition of generalization is supported by the common routine used in generative modeling consisting in early-stopping the stochastic sampling process \cite{NEURIPS2023_06abed94, Yang2021GeneralizationEO, pmlr-v145-yang22a}, with the aim of improving the quality of the examples. Consistently with \cite{NEURIPS2023_06abed94}, our analysis shows that we need a polynomial number of training data points to obtain generalization without falling into memorization.

Let us consider the exact score function measured from a data-set embedded in a linear manifold (see Section~\ref{sec:true}): we have proved in Section \ref{sec:true_emp} that the true score coincides with the empirical one for $t > t_c$. The argument around the linear manifold can be extended to a non-linear one by observing that the interesting phenomenology in DMs occur at very small times (mainly due to the data structure, see  Section~\ref{sec:generalization}), where the amplitude of the stochastic noise $\sqrt{t}$ is much smaller than the manifold curvature. A more extensive dissertation about this aspect can be found in \cite{ventura2024spectral}. 
When $F$ is a random matrix with i.i.d. standard Gaussian entries, $F^{\top} F/D$ is a Wishart matrix and its eigenvalues satisfy the Marchenko-Pastur distribution. As showed in \cite{ventura2024spectral}, the Jacobian of the empirical score function before condensation is given by
\begin{equation}
    J_{t}= \frac{1}{t}F\left[I_{D}+\frac{1}{t}F^{\top} F\right]^{-1}F^{\top}- I_{N},
    \label{eq:W_t}
\end{equation}
where we have re-absorbed the $1/D$ factor for the sake of clarity. Therefore, the spectrum of the eigenvalues of $J_t$ can be derived by a propagation of the spectrum of $F^{\top} F$ and it is
\begin{equation}
\label{eq:rho_r}
    \rho_{t}(r)=\left(1-\alpha_m\right)\delta\left(r+1\right)\theta\left[\alpha_D^{-1}-1\right]
    -\frac{\alpha_D}{2\pi}\frac{1}{r(1+r)}\sqrt{\left(r_{+}-r\right)\left(r-r_{-}\right)}\theta\left[\left(r_{+}-r\right)\left(r-r_{-}\right)\right],
\end{equation}
with $r_{\pm}(t)=-\frac{t}{\left(1\pm\frac{1}{\sqrt{\alpha_D}}\right)^{2}+t}$. The first term in $\rho_t(r)$ is a spike in $r = -1$ with mass equal to $(1-\alpha_D)$, the second term is a bulk of mass $\alpha_D$, ranging in $\left[ r_{-}(t),r_{+}(t)\right]$, and moving from $r = -1$ towards $r = 0$. 

The structure and dynamics of the eigenspectrum, composed by moving bulks of non-zeros eigenvalues towards the origin, suggest the presence of an evolving latent manifold. 
Vanishing eigenvalues are relative to tangent directions to the manifold, while non-zero ones must be associated to orthogonal directions. The score function, in fact, always points orthogonally towards the evolving manifold, since it is \textit{projecting} diffusive trajectories onto it. The final part of this transformation of the spectrum, described in detail in \cite{ventura2024spectral}, represents the consolidation of the target manifold, and it is represented by the last bulk being absorbed by the $r = 0$ spike.

We are now interested in evaluating the width of the gap forming between $r = -1$ and $r = r_{-}(t)$, i.e. the gap separating the last moving bulk and the $r = -1$ spike. We know that such gap is progressively closing when $t \rightarrow 0^+$, because the spectrum must be formed of two spikes, one of mass $(1-\alpha_D)$ in $r = -1$ and one of mass $\alpha_D$ in $r = 0$.   
We can hence find the approximate time at which the score function points towards the true target manifold by imposing such gap to equal a quantity $\delta \approx  1$. 
We call such time $t_g^{RMT}$ and it is given by
\begin{equation}
    \label{eq:tgen}
    t_{g}^{RMT}(\delta) = \left(1-\frac{1}{\sqrt{\alpha_D}}\right)^{2}\left(\frac{1-\delta}{\delta}\right).
\end{equation}
Let us compute the condition such that the score is sufficiently orthogonal to the manifold (i.e. the model generates examples that live on the data-manifold) and it has not collapsed yet. Such condition reads
\begin{equation}
    \label{eq:gen_cond}
    t_c \leq t_{g}^{RMT}(\delta).  
\end{equation}
Let us assume to be in the $D\ll\log P$ and $D\ll N$ regime where $t_c$ is given by Eq.~\eqref{eq:tcasymp}. Moreover, we choose $\delta = 1-\epsilon$ with small $\epsilon>0$. Hence, condition \eqref{eq:gen_cond} reads
\begin{equation}
\label{eq:1stc}
t_{c}\approx\frac{1}{2\alpha_D}e^{-\frac{2\alpha}{\alpha_D}}\leq   \frac{1}{2}\left(1-\frac{1}{\sqrt{\alpha_D}}\right)^{2}\left(\frac{1-\Delta}{\Delta}\right)\simeq \frac{\epsilon}{2\alpha_D},
\end{equation}
where we employed the fact that $\left(1-\alpha_D^{-1/2}\right)^2\simeq \alpha_D^{-1}$, 
when $\alpha_D\ll 1$. Eq. \eqref{eq:1stc} thus becomes $e^{\frac{2\alpha}{\alpha_D}}\geq \epsilon^{-1}$.
As a consequence, the minimum amount of data points such that the generalization condition \eqref{eq:gen_cond} is satisfied, must scale as
\begin{equation}
    P_{\text{min}}=\epsilon^{-\frac{D}{2}},
\end{equation}
which surprisingly is a function of the dimension of the manifold rather than the ambient space. 

\section{Conclusions}
We have extensively analyzed the memorization and generalization performance of a DM driven by the empirical score function, that is, the score corresponding to the noised empirical distribution, as a proxy of true or learned scores. Our main contribution is the extension the REM framework introduced by Refs. \cite{lucibello_exponential_2023,biroli_dynamical_2024} to the case of structured data living on a hidden manifold. Our study sheds light on the role of the manifold structure in learning the ground-truth distribution underneath the training set. 

Firstly, we find that empirical-score-driven DMs can both memorize and generalize a set of data points at different times. We highlighted a rich sequence of dynamical phases occurring during the reverse diffusion process that starts from $t = t_f\gg 1$ and reaches $t = 0$:
\begin{itemize}
    \item $t_c < t \leq t_o$: diffusive trajectories explore a diffusion potential which is now multistable, since data points have become local minima surrounded by basins of attraction that grow while time decreases. The typical stochastic path of the system is not trapped in one of the basins, without showing any trace of memorization. 
    \item $t_g \leq t \leq t_c$: the diffusive trajectory is now trapped in the basin of attraction, and the empirical score function points towards the closest data point. At the same time, the trajectory is also approaching the hidden data manifold. The highest proximity between the empirical distribution of the states sampled by diffusion and the ground-truth distribution of the data points is reached at $t = t_g$. This time can be interpreted as the optimal stopping time for sampling.
    \item $0 < t < t_g$: the quality of the sampled examples now deteriorates until full memorization is reached at $t = 0$.
\end{itemize}
Note that the so-called \textit{speciation time} studied in \cite{biroli_dynamical_2024, ambrogioni2024thermo}, understood as the time when the diffusive potential undergoes a spontaneous symmetry breaking into multiple ergodic components that are representative of the data classes, has not been analyzed in our paper since our data model does not have clear class separation. We refer the reader to \cite{macris} for the study of the speciation time under the manifold hypothesis.   

Surprisingly, the best degree of generalization is reached inside the memorization phase of the model, while the score function drives the model towards the closest attractor.
The dynamical picture of the DM reported above is deformed by the presence of structure in the data, as it emerged from our analysis. Specifically, when $\alpha$ is fixed and $\alpha_D \rightarrow 0$:
\begin{enumerate}
    \item Even though the onset time exponentially decreases, the distance between $t_o$ and the condensation time increases until reaching a constant plateau.  
    \item The collapse time $t_c$ shrinks towards $t = 0$, and the empirical-score-drive DM tends to the exact model, hence reducing the volume of the memorization phase of the model. This result is consistent with the very recent result obtained by \cite{macris} in the matter of variance-preserving DMs.
    \item The generalization time $t_g$ also moves towards $t = 0$, yet faster than $t_c$. 
\end{enumerate}
In light of point (3) we conclude that DMs benefit from highly structured data, even when $\alpha_D$ has not completely vanished, since the model can be basically stopped at $t \simeq 0$ and obtain a good degree of generalization, as one would obtain through a neural-network-trained model. 

As an alternative to this definition of generalization, we use a combination of the REM formalism and Random Matrix Theory (RMT) to provide the reader with the minimal number of training data point to build the empirical score function in such a way that the DM is capable of sampling from the manifold 
with a minimal KL divergence.
We find that the size of the data set needs to scale exponentially with the latent dimension of the data, instead of the visible dimension, mitigating the curse of dimensionality that affects learning in generative models \cite{yarotzky, cybenko}. 

\section{Acknowledgments}
CL and EV acknowledge European Union - Next Generation EU funds, component M4.C2, investment 1.1 - CUP J53D23001330001. MM work has been
supported by the PNRR-PE-AI FAIR project funded by the NextGeneration EU program.

\bibliography{main}

\newpage
\appendix
\section{Collapse Time for Homogeneous Gaussian Data}

When the data points live a linear manifold we can
consider the basis in which the manifold has diagonal covariance matrix
$\Sigma$ with elements $\sigma_{i}^{2}$ (distributed according to
the Marchenko-Pastur distribution). 

In order to investigate the scaling, let's simplify and assume that
$D$ dimensions have variance $\sigma_{i}^{2}=\sigma^{2}$ and $N-D$
have variance $\sigma_{i}^{2}=0$. We have

\begin{equation}
\zeta_{t}(\lambda)=-\frac{1}{2}\alpha_{D}\log(1+\frac{\lambda}{\alpha_{D} t}\sigma^{2})-\frac{\lambda}{2}\alpha_{D}\frac{t\alpha_{D}+\sigma^{2}}{t\alpha_{D}+\lambda\sigma^{2}}-\frac{\lambda}{2}(1-\alpha_{D})
\end{equation}

We can find the collapse time from the condition

\begin{align}
\zeta_{t_{c}}(1)+\alpha & =-\frac{1}{2}
\end{align}

which implies

\begin{align}
-\alpha_{D}\log(1+\frac{\sigma^{2}}{\alpha_{D}t})-1+2\alpha & =-1
\end{align}

The solution is 
\begin{align}
t_{c} & =\frac{\sigma^{2}N/D}{e^{2\log P/D}-1}.
\end{align}

It results that the collapse time depends on the manifold dimension
and the number of hidden points.
If we consider the limit of $D\ll\log P$ and $D\ll N$ we have
\begin{equation}
t_{c}\approx\sigma^{2}\frac{N}{D}e^{-\frac{2\log P}{D}}
\end{equation}
which goes to zero exponentially fast.

\subsection{Variance Preserving Case \label{app:vp}}

If instead we consider the variance preserving framework where $x=\xi^{1}e^{-t}+\omega\sqrt{\Delta_{t}}$, $\Delta_{t}=1-e^{-2t}$, the cumulant generating function reads 

\begin{equation}
\zeta_{t}(\lambda)=-\frac{1}{2}\alpha_{D}\log(1+\frac{\lambda e^{-2t}}{\Delta_{t}}\sigma^{2})-\frac{\lambda}{2}\alpha_{D}\frac{\Delta_{t}+\sigma^{2}e^{-2t}}{\Delta_{t}+\lambda\sigma^{2}e^{-2t}}-\frac{\lambda}{2}(1-\alpha_{D})
\end{equation}

We can find the collapse time from the condition

\begin{align}
\zeta_{t_{c}}(1)+\alpha & =-\frac{1}{2}
\end{align}

which implies

\begin{align}
-\alpha_{D}\log(1+\frac{e^{-2t}}{1-e^{-2t}}\sigma^{2})-1+2\alpha & =-1
\end{align}

The solution is 
\begin{align}
t_{c} & =\frac{1}{2}\log\left(1+\frac{\sigma^{2}}{e^{2\alpha/\alpha_{D}}-1}\right)\\
 & =\frac{1}{2}\log\left(1+\frac{\sigma^{2}}{e^{\frac{2\log P}{D}}-1}\right).
\end{align}
Such expression for the collapse time has been also recently found by \cite{macris}, and it can be easily compared with the one found in \cite{biroli_dynamical_2024} for $\alpha_D=1$, i.e. the homogeneous Gaussian case.

\section{Condensation Time: Computation of the Generating function}
\subsection{Linear case}\label{appendix:zeta-linear}

In the variance exploding case we have
\begin{align}
&\zeta_{t}(\lambda)= \lim_{N\to\infty}\frac{1}{N}\mathbb{E}_{F,z^{1},\omega}\log\mathbb{E}_{z^{2}}e^{-\frac{\lambda}{2t}\lVert\left(\frac{F}{\sqrt{D}}z^{2}-\frac{F}{\sqrt{D}}z^{1}\right)+\omega\sqrt{t}\rVert^{2}}\\
&=  \lim_{N\to\infty}\frac{1}{N}\mathbb{E}_{F,z^{1},\omega}\log\int\frac{dz^{2}}{2\pi}\ e^{-\frac{1}{2}z^{2}(I+\frac{\lambda}{t}\frac{F^{T}F}{D})z^{2}+\frac{\lambda}{t}z^{2}(\frac{F^{T}F}{D}z^{1}-\frac{F^T}{\sqrt{D}}\omega\sqrt{t})-\frac{\lambda}{2t}\lVert-\frac{F}{\sqrt{D}}z^{1}+\omega\sqrt{t}\rVert^{2}}\\
&=  \lim_{N\to\infty}\frac{1}{N}\mathbb{E}_{F,z^{1},\omega}\Bigg[ -\frac{1}{2}\log\det(I+\frac{\lambda}{t}\frac{F^{T}F}{D}) \nonumber \\
&+\frac{1}{2}\frac{\lambda^{2}}{t^{2}}(\frac{F^{T}F}{D}z^{1}-\frac{F^T}{\sqrt{D}}\omega\sqrt{t})^{T}(I+\frac{\lambda}{t}\frac{F^TF}{D})^{-1}(\frac{F^{T}F}{D}z^{1}-\frac{F^T}{\sqrt{D}}\omega\sqrt{t})\nonumber \\
&-\frac{\lambda}{2t}\lVert-\frac{F}{\sqrt{D}}z^{1}+\omega\sqrt{t}\rVert^{2}\Bigg]\\
&=  \lim_{N\to\infty}\frac{1}{N}\mathbb{E}_{F,z^{1},\omega}\Bigg[ -\frac{1}{2}\log\det(I+\frac{\lambda}{t}\frac{F^{T}F}{D})+\frac{\lambda^{2}}{2t^{2}}(\frac{F}{\sqrt{D}}z^{1})^{T}\frac{F}{\sqrt{D}}(I+\frac{\lambda}{t}\frac{F^{T}F}{D})^{-1}\frac{F^{T}}{\sqrt{D}}(\frac{F}{\sqrt{D}}z^{1}) \nonumber\\
 &-\frac{\lambda}{2t}\lVert-\frac{F}{\sqrt{D}}z^{1}\lVert^{2} +\frac{\lambda^{2}}{2t^{2}}(\frac{F^{T}}{\sqrt{D}}\omega\sqrt{t})^{T}(I+\frac{\lambda}{t}\frac{F^{T}F}{D})^{-1}(\frac{F^{T}}{\sqrt{D}}\omega\sqrt{t})-\frac{\lambda}{2t}\lVert\omega\sqrt{t}\rVert^{2}\Bigg]
\end{align}
Now with a rotation we can position in the basis of the eigenvectors
of $\frac{F^{\top}F}{N}$, with eigenvalues $\sigma^{2}_k$. 
\begin{align}
 & =\lim_{N\to\infty}\frac{1}{N}\sum_{i}^{N}\left[-\frac{\lambda}{2}\right]+\frac{1}{N}\sum_{k}^{D}\left[-\frac{1}{2}\log(1+\frac{\lambda}{\alpha_D t}\sigma_{k}^{2})+\frac{\lambda^{2}}{2 \alpha_D^2 t^{2}}(\frac{\sigma_{k}^{4}}{1+\frac{\lambda}{\alpha_D t}\sigma_{k}^{2}})-\frac{\lambda}{2\alpha_D t}\sigma_{k}^{2}+\frac{\lambda^{2}}{2\alpha_D t^{2}}(\frac{t\sigma_{k}^{2}}{1+\frac{\lambda}{\alpha_D t}\sigma_{k}^{2}})\right]\\
 & =-\frac{\lambda}{2} + \lim_{N\to\infty}\frac{1}{N}\sum_{k}\left[-\frac{1}{2}\log(1+\frac{\lambda}{\alpha_D t}\sigma_{k}^{2})+\frac{\lambda^{2}}{2\alpha_D t}(\frac{\sigma_{k}^{4}}{\alpha_D t+\lambda\sigma_{k}^{2}})-\frac{\lambda}{2\alpha_D t}\sigma_{k}^{2}+\frac{\lambda^{2}}{2t}(\frac{t\sigma_{k}^{2}}{\alpha_D t+\lambda\sigma_{k}^{2}})\right]\\
 & =-\frac{\lambda}{2} + \lim_{N\to\infty}\frac{1}{N}\sum_{k}\left[-\frac{1}{2}\log(1+\frac{\lambda}{\alpha_D t}\sigma_{k}^{2})\right]
\end{align}
Here we have assumed that $\alpha_{D}<1$. Taking the limit $N\to\infty$ the sum becomes an integration over the distribution $\nu$ of $\sigma^{2}$, which is the bulk of a Marchenko-Pastur distribution

\begin{equation}
\zeta_{t}(\lambda)= - \frac{\lambda}{2} -\frac{\alpha_D}{2}\int\nu_{\alpha_{D}}(d\sigma^{2})\log\left(1+\frac{\lambda \sigma^{2}}{\alpha_D t}\right)
\end{equation}
with
\begin{align}
d\nu_{\gamma}(x) & =\frac{1}{2\pi}\frac{\sqrt{(\gamma^{+}-x)(\gamma^{-}-x)}}{\gamma x}\mathbb{I}\left(x\in[\gamma^{-},\gamma^{+}]\right)\\
\gamma^{\pm} & =(1\pm\sqrt{\gamma})^{2}
\end{align}
If we compute everything at $\lambda=1$ this becomes
\begin{equation}
\zeta_{t}(1)=-\frac{1}{2}\int\nu_{\alpha_{D}}(d\sigma^{2})\left[\log(1+\frac{\sigma^{2}}{\alpha_D t})\right]-\frac{1}{2}
\label{eq:zeta-mp}
\end{equation}
Taking the derivative
\begin{align}
\zeta_t'(\lambda) & =-\frac{\alpha_D}{2} \int\nu_{\alpha_{D}}(d\sigma^{2})\frac{\sigma^{2}}{\alpha_D t+\lambda\sigma^{2}}\\
\zeta_t'(1) & =-\frac{1}{2}
\end{align}
We can also use replica theory, which will be necessary in the non-linear case, and compare the results. The replicated $\zeta_t$ reads
\begin{equation}
\mathbb{E}\mathcal{Z}^{n}=\mathbb{E}_{F,\omega}\mathbb{E}_{z^{0:n}}\ e^{-\frac{\lambda}{2t}\sum_{a'}\lVert\frac{Fz^{0}}{\sqrt{D}}-\frac{Fz^{a'}}{\sqrt{D}}\lVert^{2}-\frac{\lambda\omega}{t}\sum_{a'}\left(\frac{Fz^{0}}{\sqrt{D}}-\frac{Fz^{a'}}{\sqrt{D}}\right)-\frac{\lambda}{2}\rVert\omega\lVert^{2}}
\end{equation}

\begin{align}
\mathbb{E}\mathcal{Z}^{n} &=\mathbb{E}_{F,\omega}\mathbb{E}_{z^{0:n}}\ e^{-\frac{\lambda}{2t}\sum_{a'}\lVert\frac{Fz^{0}}{\sqrt{D}}-\frac{Fz^{a'}}{\sqrt{D}}\lVert^{2}-\frac{\lambda\omega}{t}\sum_{a'}\left(\frac{Fz^{0}}{\sqrt{D}}-\frac{Fz^{a'}}{\sqrt{D}}\right)-\frac{\lambda}{2}\rVert\omega\lVert^{2}}\\
 & =\mathbb{E}_{F,\omega}\mathbb{E}_{z^{0:n}}\int\frac{d\hat{u}du}{2\pi}e^{-\frac{\lambda}{2t}\sum_{i}\sum_{a'}\left(u_{i}^{0}-u_{i}^{a'}\right)^2 -\frac{\lambda}{2}\sum_i \sum_{a'} \omega_i(u_i^0 - u_i^{a'}) -\frac{\lambda}{2}\sum_i \omega_i^2} e^{-i\sum_{i}\sum_{a=0}^{n}\hat{u}_{i}^{a}u_{i}^{a}+\sum_{a}\frac{i}{\sqrt{D}}\sum_{ik}\hat{u}_{i}^{a}F_{ik}z_{k}^{a}}\\
 & =\mathbb{E}_{\omega}\mathbb{E}_{z^{0:n}}\int\frac{d\hat{u}du}{2\pi}e^{-\frac{\lambda}{2t}\sum_{i}\sum_{a'}\left(u_{i}^{0}-u_{i}^{a'}\right)^2 -\frac{\lambda}{2}\sum_i \sum_{a'} \omega_i(u_i^0 - u_i^{a'}) -\frac{\lambda}{2}\sum_i \omega_i^2}e^{-i\sum_{i}\sum_{a=0}^{n}\hat{u}_{i}^{a}u_{i}^{a}-\frac{1}{2D}\sum_{ik}\left(\sum_{a}\hat{u}_{i}^{a}z_{k}^{a}\right)^{2}}\\
 & =\mathbb{E}_{\omega}\mathbb{E}_{z^{0:n}}\int\frac{d\hat{u}du}{2\pi}e^{-\frac{\lambda}{2t}\sum_{i}\sum_{a'}\left(u_{i}^{0}-u_{i}^{a'}\right)^2 -\frac{\lambda}{2}\sum_i \sum_{a'} \omega_i(u_i^0 - u_i^{a'}) -\frac{\lambda}{2}\sum_i \omega_i^2}e^{-i\sum_{i}\sum_{a=0}^{n}\hat{u}_{i}^{a}u_{i}^{a}-\frac{1}{2D}\sum_{ab}\left(\sum_{i}\hat{u}_{i}^{a}\hat{u}_{i}^{b}\right)\left(\sum_{k}z_{k}^{a}z_{k}^{b}\right)}\\
 & =\int dq\,d\hat{q}\ e^{nN\mathcal{\phi}_{\lambda}(q,\hat{q})}
\end{align}
with the overlaps defined as
\begin{equation}
q_{ab}=\frac{1}{D}\sum_{k}z_{k}^{a}z_{k}^{b}
\end{equation}
so that we can write the replicated action
\begin{equation}
\zeta_t(\lambda, t; q,\hat{q})=-\frac{1}{2n}\frac{D}{N}\sum_{ab=0}^{n}q_{ab}\hat{q}_{ab}+\frac{D}{N}G_{S}(\hat{q})+G_{E}(\lambda, t; q)
\end{equation}
with
\begin{align}
G_{S}(\hat{q}) & =\frac{1}{n}\log\mathbb{E}_{z^{0:n}}\ e^{\frac{1}{2}\sum_{ab}\hat{q}_{ab}z^{a}z^{b}}\\
G_{E}(\lambda,t;q) & =\frac{1}{n}\log\int D\omega\int\prod_{a=0}^{n}\frac{d\hat{u}_{a}du_{a}}{2\pi}e^{-\frac{\lambda}{2t}\sum_{a'}(u^{0}-u^{a'}+\omega\sqrt{t})^{2}-i\sum_{a=0}^{n}\hat{u}^{a}u^{a}-\frac{1}{2}\sum_{ab}\hat{u}^{a}\hat{u}^{b}q_{ab}}
\end{align}
Using the replica symmetric ansatz 
\begin{equation}
q_{ab}=\left(\begin{array}{cccc}
1 & m & \ldots & m\\
m & q_{d} &  & q_{0}\\
\vdots &  & \ddots\\
m & q_{0} &  & q_{d}
\end{array}\right);\ \ \ \hat{q}_{ab}=\left(\begin{array}{cccc}
0 & \hat{m} & \ldots & \hat{m}\\
\hat{m} & \hat{q}_{d} &  & \hat{q}_{0}\\
\vdots &  & \ddots\\
\hat{m} & \hat{q}_{0} &  & \hat{q}_{d}
\end{array}\right)\label{eq:rs-ansatz}
\end{equation}
we find
\begin{equation}
\zeta_t(\lambda;q_{d},q_{0},m,\hat{q}_{d},\hat{q}_{0},\hat{m})=-\alpha_{D}m\hat{m}-\frac{\alpha_{D}}{2}(q_{d}\hat{q}_{d}-q_{0}\hat{q}_{0})+\alpha_{D}G_{S}(\hat{q}_{d},\hat{q}_{0},\hat{m})+G_{E}(\lambda, t; q_{d},q_{0},m)
\end{equation}
with
\begin{align}
G_{S}(\hat{q}_{d},\hat{q}_{0},\hat{m}) & =-\frac{1}{2}\log\left(1-\hat{q}_{d}+\hat{q}_{0}\right)+\frac{1}{2}\frac{\hat{m}^{2}+\hat{q}_{0}}{1-\hat{q}_{d}+\hat{q}_{0}}
\end{align}
and for the energetic term
{\small
\begin{align}
G_{E} & =\int D\omega\int D\gamma\int Du^{0}\log\left(\int Du\ e^{-\frac{\lambda}{2t}\left(u^{0}-\sqrt{q_{d}-q_{0}}u-mu^{0}+\sqrt{q_{0}-m^{2}}\gamma\right)^{2}-\frac{\lambda\omega}{\sqrt{t}}\left(u^{0}-\sqrt{q_{d}-q_{0}}u-mu^{0}+\sqrt{q_{0}-m^{2}}\gamma\right)-\frac{\lambda}{2}\omega{}^{2}}\right)\\
 & =\int D\omega\int D\gamma\int Du^{0}\Big[-\frac{\lambda}{2t}\left((1-m)u^{0}+\sqrt{q_{0}-m^{2}}\gamma\right)^{2}-\frac{\lambda\omega}{\sqrt{t}}\left((1-m)u^{0}+\sqrt{q_{0}-m^{2}}\gamma\right)-\frac{\lambda}{2}\omega{}^{2}
 \nonumber\\
 & -\frac{1}{2}\log\left(1+\frac{\lambda(q_{d}-q_{0})}{t}\right)\nonumber\\
 &+\frac{1}{2}\left(1+\frac{\lambda(q_{d}-q_{0})}{t}\right)^{-1}\left(\frac{\lambda}{t}\sqrt{q_{d}-q_{0}}\left((1-m)u^{0}+\sqrt{q_{0}-m^{2}}\gamma\right)+\frac{\lambda\omega}{t}\sqrt{q_{d}-q_{0}}\right)^{2}\Big]\\
 & =-\frac{1}{2}\left(\lambda+\log\left(1+\frac{\lambda(q_{d}-q_{0})}{t}\right)+\frac{\lambda}{t}(1-2m+q_{0})-\frac{\lambda^{2}(q_{d}-q_{0})(1-2m+q_{0}+t)}{t^{2}\left(1+\frac{\lambda(q_{d}-q_{0})}{t}\right)}\right)
\end{align}
}
In order to compute the typical value of $\zeta_t(\lambda)$ to employ for solving the collapse condition, we need to derive the following saddle point equations 
\begin{align}
\frac{\partial\zeta_t}{\partial\hat{q}_{d}} & =0 & q_{d} & =\frac{1}{1-\hat{q}_{d}+\hat{q}_{0}}+\frac{\hat{m}^{2}+\hat{q}_{0}}{\left(1-\hat{q}_{d}+\hat{q}_{0}\right)^{2}}\\
\frac{\partial\zeta_t}{\partial\hat{q}_{0}} & =0 & q_{0} & =\frac{1}{1-\hat{q}_{d}+\hat{q}_{0}}+\frac{\hat{m}^{2}+\hat{q}_{d}-1}{\left(1-\hat{q}_{d}+\hat{q}_{0}\right)^{2}}\\
\frac{\partial\zeta_t}{\partial\hat{m}} & =0 & m & =\frac{\hat{m}}{1-\hat{q}_{d}+\hat{q}_{0}}\\
\frac{\partial\zeta_t}{\partial q_{d}} & =0 & \hat{q}_{d} & =\frac{2}{\alpha_{D}}\left(-\frac{1}{2}\right)\frac{\lambda\left(t+(-1-2m+q_{0}+t)\lambda\right)}{\left(t+(q-q_{0})\lambda\right)^{2}}\\
\frac{\partial\zeta_t}{\partial q_{0}} & =0 & \hat{q}_{0} & =-\frac{2}{\alpha_{D}}\left(-\frac{1}{2}\right)\frac{(1-2m+q_{0}+t)\lambda^{2}}{\left(t+(q-q_{0})\lambda\right)^{2}}\\
\frac{\partial\zeta_t}{\partial m} & =0 & \hat{m} & =\frac{1}{\alpha_{D}}\left(-\frac{1}{2}\right)\left(-\frac{2\lambda}{t}+\frac{2(q-q_{0})\lambda^{2}}{t^{2}\left(1+\frac{(q-q_{0})\lambda}{t}\right)}\right).
\end{align}
By solving these saddle point equations we recover perfect agreement
with Eq.~\eqref{eq:zeta-mp}.

\subsection{Non-linear case}\label{appendix:zeta-non-lin-1}

In the non-linear case, the replicated partition function reads
\begin{align}
\mathbb{E}\mathcal{Z}^{n} & =\mathbb{E}_{F,\omega}\mathbb{E}_{z^{0:n}}\ e^{-\frac{\lambda}{2t}\sum_{a'}\lVert g\left(\frac{Fz^{0}}{\sqrt{D}}\right)-g\left(\frac{Fz^{a'}}{\sqrt{D}}\right) + \omega \sqrt{t}\lVert^{2}}\\
 & =\mathbb{E}_{F,\omega}\mathbb{E}_{z^{0:n}}\int\frac{d\hat{u}du}{2\pi}e^{-\frac{\lambda}{2t}\sum_{i}\sum_{a'}\left(g\left(u_{i}^{0}\right)-g\left(u_{i}^{a'}\right)\right) + \omega_i \sqrt{t})^{2}}e^{-i\sum_{i}\sum_{a=0}^{n}\hat{u}_{i}^{a}u_{i}^{a}+\sum_{a}\frac{i}{\sqrt{D}}\sum_{ik}\hat{u}_{i}^{a}F_{ik}z_{k}^{a}}\\
 & =\mathbb{E}_{\omega}\mathbb{E}_{z^{0:n}}\int\frac{d\hat{u}du}{2\pi}e^{-\frac{\lambda}{2t}\sum_{i}\sum_{a'}\left(g\left(u_{i}^{0}\right)-g\left(u_{i}^{a'}\right) +\omega_i \sqrt{t}\right)^{2}}e^{-i\sum_{i}\sum_{a=0}^{n}\hat{u}_{i}^{a}u_{i}^{a}-\frac{1}{2D}\sum_{ik}\left(\sum_{a}\hat{u}_{i}^{a}z_{k}^{a}\right)^{2}}\\
 & =\mathbb{E}_{\omega}\mathbb{E}_{z^{0:n}}\int\frac{d\hat{u}du}{2\pi}e^{-\frac{\lambda}{2t}\sum_{i}\sum_{a'}\left(g\left(u_{i}^{0}\right)-g\left(u_{i}^{a'}\right)+\omega_i \sqrt{t}\right)^{2}}e^{-i\sum_{i}\sum_{a=0}^{n}\hat{u}_{i}^{a}u_{i}^{a}-\frac{1}{2D}\sum_{ab}\left(\sum_{i}\hat{u}_{i}^{a}\hat{u}_{i}^{b}\right)\left(\sum_{k}z_{k}^{a}z_{k}^{b}\right)}\\
 & =\int dq\,d\hat{q}\ e^{nN\mathcal{\phi}_{\lambda}(q,\hat{q})}
\end{align}
with the overlaps defined as
\begin{equation}
q_{ab}=\frac{1}{D}\sum_{k}z_{k}^{a}z_{k}^{b}
\end{equation}
so that we can write the replicated action
\begin{equation}
\zeta_t(q,\hat{q})=-\frac{1}{2n}\frac{D}{N}\sum_{ab=0}^{n}q_{ab}\hat{q}_{ab}+\frac{D}{N}G_{S}(\hat{q})+G_{E}(q)
\end{equation}
with
\begin{align}
G_{S} & =\frac{1}{n}\log\mathbb{E}_{z^{0:n}}\ e^{\frac{1}{2}\sum_{ab}\hat{q}_{ab}z^{a}z^{b}}\\
G_{E} & =\frac{1}{n}\log\int D\omega\int\prod_{a=0}^{n}\frac{d\hat{u}_{a}du_{a}}{2\pi}e^{-\frac{\lambda}{2t}\sum_{a'}(g\left(u^{0}\right)-g\left(u^{a'}\right)+\omega\sqrt{t})^{2}-i\sum_{a=0}^{n}\hat{u}^{a}u^{a}-\frac{1}{2}\sum_{ab}\hat{u}^{a}\hat{u}^{b}q_{ab}}
\end{align}
We invoke the replica symmetric ansatz as performed in the linear case (see Appendix \ref{appendix:zeta-linear}) and obtain the same expression for $\zeta_t(q_{d},q_{0},m,\hat{q}_{d},\hat{q}_{0},\hat{m})$ with a different energetic term, due to the non-linearity, that reads





\begin{align}
G_{E} & =\frac{1}{n}\log\int D\omega\int\prod_{a=0}^{n}\frac{d\hat{u}_{a}du_{a}}{2\pi}e^{-\frac{\lambda}{2t}\sum_{a'}(g\left(u^{0}\right)-g\left(u^{a'}\right)+\omega\sqrt{t})^{2}-i\sum_{a=0}^{n}\hat{u}^{a}u^{a}-\frac{1}{2}\sum_{ab}\hat{u}^{a}\hat{u}^{b}q_{ab}}\\
 & =\frac{1}{n}\log\int D\omega\int\frac{du^{0}d\hat{u}^{0}}{2\pi}\prod_{a'=1}^{n}\frac{du^{a'}d\hat{u}^{a'}}{2\pi}\ e^{-\frac{\lambda}{2t}\sum_{a'}(g\left(u^{0}\right)-g\left(u^{a'}\right)+\omega\sqrt{t})^{2}}\nonumber\\
 & \times e^{-\sum_{a'}i\hat{u}^{a'}u^{a'}-i\hat{u}^{0}u^{0}-\frac{1}{2}\left(\hat{u}^{0}\right)^{2}-m\hat{u}^{0}\sum_{a'}\hat{u}^{a'}-\frac{1}{2}(q_{d}-q_{0})\sum_{a'}\left(\hat{u}^{a'}\right)^{2}-\frac{1}{2}q_{0}\left(\sum_{a'}\hat{u}^{a'}\right)^{2}}\\
 & =\frac{1}{n}\log\int D\omega\int\frac{du^{0}}{\sqrt{2\pi}}\prod_{a'=1}^{n}\frac{du^{a'}d\hat{u}^{a'}}{2\pi}\ e^{\frac{1}{2}\left(m\sum_{a'}\hat{u}^{a'}+iu^{0}\right)^{2}}e^{-\frac{\lambda}{2t}\sum_{a'}(g\left(u^{0}\right)-g\left(u^{a'}\right)+\omega\sqrt{t})^{2}}\nonumber\\
 & \times e^{-\sum_{a'}i\hat{u}^{a'}u^{a'}-\frac{1}{2}(q_{d}-q_{0})\sum_{a'}\left(\hat{u}^{a'}\right)^{2}-\frac{1}{2}q_{0}\left(\sum_{a'}\hat{u}^{a'}\right)^{2}}\\
 & =\frac{1}{n}\log\int D\omega\int D\gamma\int\frac{du^{0}}{\sqrt{2\pi}}\;e^{-\frac{1}{2}\left(u^{0}\right)^{2}}\int\prod_{a'=1}^{n}\frac{du^{a'}d\hat{u}^{a'}}{2\pi}\ e^{-\frac{\lambda}{2t}\sum_{a'}(g\left(u^{0}\right)-g\left(u^{a'}\right)+\omega\sqrt{t})^{2}}\nonumber\\
 & \times e^{-i\sum_{a'}\hat{u}^{a'}\left(u^{a'}-mu_{0}+\sqrt{q_{0}-m^{2}}\gamma\right)-\frac{1}{2}(q_{d}-q_{0})\sum_{a'}\left(\hat{u}^{a'}\right)^{2}}\\
 & =\frac{1}{n}\log\int D\omega\int D\gamma\int Du^{0}\left(\int\frac{du}{\sqrt{2\pi}}\ e^{-\frac{\lambda}{2t}(g\left(u^{0}\right)-g\left(u\right)+\omega\sqrt{t})^{2}}\frac{1}{\sqrt{q_{d}-q_{0}}}e^{-\frac{1}{2(q_{d}-q_{0})}\left(u-mu^{0}+\sqrt{q_{0}-m^{2}}\gamma\right)^{2}}\right)^{n}\\
 & =\int D\omega\int D\gamma\int Du^{0}\log\left(\int Du\ e^{-\frac{\lambda}{2t}\left(g\left(u^{0}\right)-g\left(\sqrt{q_{d}-q_{0}}u+mu^{0}-\sqrt{q_{0}-m^{2}}\gamma\right)+\sqrt{t}\omega\right)^{2}}\right).
\end{align}
We can then take derivatives to obtain the saddle point equations, which will of course depend on the choice of the non-linearity $g$, and solve them numerically, to obtain the typical $\zeta_t(\lambda)$. At this point one can solve the collapse condition and recover the memorization time as in Fig. \ref{fig:tc_nl-scaling}. 

\section{Equivalence between Collapse and Condensation}\label{appendix:equivalence}

In order to establish that the condensation and collapse phenomena
happen at the same time, $t_{c}=t_{cond}$, we would therefore need
to prove that
\begin{equation}
\zeta'_{t_{c}}(1)=-\frac{1}{2}.
\end{equation}
We consider a typical data to be a diffused version of one of the
starting training points
\begin{equation}
x_{t}=\xi^{1}+\sqrt{t}\omega.
\end{equation}
Notice that here we use the variance exploding diffusion process
for homogeneity with the rest of the paper, but this analysis does
not depend on the diffusion protocol, as long as we consider a typical
point.

We now write $\zeta_t(\lambda)$ as
\begin{equation}
\zeta_t(\lambda)=\mathbb{E}_{\xi^{1}}\mathbb{E}_{p(x_{t}|\xi^{1})}\log\mathbb{E}_{\xi}p_{\lambda}(x_{t}|\xi)
\end{equation}
where the data points come form a prior distribution, $\xi^{1},\xi\sim p(\xi)$, and the likelihood has the form

\begin{equation}
p_{\lambda}(x_{t}|\xi)\propto e^{-\frac{\lambda}{2 t}\|x_{t}-\xi\|^{2}}.
\end{equation}
Then we compute $\zeta_t'(\lambda)$ taking the derivative
\begin{align}
\partial_{\lambda}\log\mathbb{E}_{\xi}p_{\lambda}(x_{t}|\xi) & =\frac{\int-\frac{\|x_{t}-\xi\|^{2}}{2 t}p_{\lambda}(x_{t}|\xi)p(\xi)\,d\xi}{\int p_{\lambda}(x_{t}|\xi)p(\xi)\,d\xi}\\
 & =\int-\frac{\|x_{t}-\xi\|^{2}}{2t}p_{\lambda}(\xi|x_{t})\,d\xi
\end{align}
so we can write this quantity as an average with respect to the posterior
distribution $p(\xi|x_{t})$, which we will indicate with $\langle\cdot\rangle_{\xi|x}$.
Substituting $\lambda=1$ and applying the Nishimori condition \cite{nishimori} we
finally obtain
\begin{align}
\zeta_t'(1) & =\mathbb{E}_{\xi^{1}}\left[\mathbb{E}_{x|\xi^{1}}\left[\langle-\frac{\|x_{t}-\xi\|^{2}}{2 t}\rangle_{\xi|x}\right]\right]\\
 & =\mathbb{E}_{\xi^{1}}\left[\mathbb{E}_{x|\xi^{1}}\left[-\frac{\|x_{t}-\xi^{1}\|^{2}}{2t}\right]\right]\\
 & =-\frac{1}{2}.
\end{align}

\section{Onset Time: Computation of the Generating function}
\label{sec:ot}
\subsection{Linear case}
\label{sec:ot_lin}

As explained in Section \ref{sec:REM}, we need to compute the cumulant generating function as
\begin{align}
\zeta_{t}(\lambda)= & \lim_{N\to\infty}\frac{1}{N}\mathbb{E}_{F,z^{0}}\log\mathbb{E}_{z}e^{-\frac{\lambda}{2t}\lVert\left(\frac{Fz}{\sqrt{D}}-\frac{Fz^{0}}{\sqrt{D}}\right)\rVert^{2}}\\
= & \lim_{N\to\infty}\frac{1}{N}\mathbb{E}_{F,z^{0}}\log\int\frac{dz}{\sqrt{2\pi}}\ e^{-\frac{1}{2}z(I_D+\frac{\lambda}{t}\frac{F^{T}F}{D})z+\frac{\lambda}{t}z(\frac{F^{T}F}{D}z^{0})-\frac{\lambda}{2t}\lVert-\frac{Fz^{0}}{\sqrt{D}}\rVert^{2}}\\
= & \lim_{N\to\infty}\frac{1}{N}\mathbb{E}_{F,z^{0}}\Bigg[ -\frac{1}{2}\log\det\left(I_D+\frac{\lambda}{t}\frac{F^{T}F}{D}\right)\nonumber \\
&+\frac{1}{2}\frac{\lambda^{2}}{t^{2}}\left(\frac{F^{T}F}{D}z^{0}\right)^{T}\left(I_D+\frac{\lambda}{t}\frac{F^{T}F}{D}\right)^{-1}\left(\frac{F^{T}F}{D}z^{0}\right) -\frac{\lambda}{2t}\lVert \frac{Fz^{0}}{\sqrt{D}}\rVert^{2}\Bigg].
\end{align}
Now with a rotation we can position in the basis of the eigenvectors
of $\frac{F^{\top}F}{N}$, with eigenvalues $\sigma_k^{2}$ 
\begin{align}
 & =\lim_{N\to\infty}\frac{1}{N}\sum_{k}^{D}\left[-\frac{1}{2}\log\left(1+\frac{\lambda}{\alpha_D t}\sigma_{k}^{2}\right)+\frac{\lambda^{2}}{2\alpha_D^2 t^{2}}\left(\frac{\sigma_{k}^{4}}{1+\frac{\lambda}{\alpha_D t}\sigma_{k}^{2}}\right)-\frac{\lambda}{2\alpha_D t}\sigma_{k}^{2}\right] \\
 &=\lim_{N\to\infty}\frac{1}{N}\sum_{k}\left[-\frac{1}{2}\log\left(1+\frac{\lambda}{\alpha_D t}\sigma_{k}^{2}\right)-\frac{\lambda}{2}\frac{\sigma_{k}^{2}}{\alpha_D t+\lambda\sigma_{k}^{2}}\right].
\end{align}
Here we have assumed that $\alpha_{D}<1$. Replacing with the law $\nu$ for the bulk of the Marchenko-Pastur distribution we have
\begin{equation}
\zeta_{t}(\lambda)=-\frac{\alpha_D}{2}\int\nu_{\alpha_{D}}(d\sigma^{2})\left[\log\left(1+\frac{\lambda \sigma^{2}}{\alpha_D t}\right)+\frac{\lambda\sigma^{2}}{\alpha_D t+\lambda\sigma^{2}}\right].
\end{equation}
This expression of $\zeta_t$ at $\lambda=1$ is then used to obtain $\phi(\alpha, t)$.

\subsection{Non-linear case}
\label{sec:ot_nlin}

In case of non-linear functions that define the manifold, we are going to employ the replica method to compute the REM free-energy, as we performed for the condensation time. First we need to compute the cumulant generating function as
\begin{align}
    \zeta_{t}(\lambda) &= \lim_{N\to \infty} \frac{1}{N} \mathbb{E}_x\log \mathbb{E}_{\xi}[e^{-\frac{\lambda}{2t}\|x-\xi\|^2}] \\
    & = \lim_{N\to \infty} \frac{1}{N} \mathbb{E}_{F,z_0} \log \mathbb{E}_{z}[e^{-\frac{\lambda}{2t}\|g(\frac{Fz_0}{\sqrt{D}})-g(\frac{Fz}{\sqrt{D}})\|^2}] \\
    & = \lim_{N\to \infty} \frac{1}{N} \mathbb{E}_{F,z_a} [e^{-\frac{\lambda}{2t}\sum_{a'}\|g(\frac{Fz_0}{\sqrt{D}})-g(\frac{Fz_{a'}}{\sqrt{D}})\|^2}].
\end{align}
Using the replica symmetric ansatz we obtain
\begin{equation}
\zeta_{t}(\lambda; q_{d},q_{0},m,\hat{q}_{d},\hat{q}_{0},\hat{m})=-\alpha_{D}m\hat{m}-\frac{\alpha_{D}}{2}(q_{d}\hat{q}_{d}-q_{0}\hat{q}_{0})+\alpha_{D}G_{S}(\hat{q}_{d},\hat{q}_{0},\hat{m})+G_{E}(\lambda, t; q_{d},q_{0},m)
\end{equation}
with
\begin{equation}
    G_{S}(\hat{q}_{d},\hat{q}_{0},\hat{m}) =-\frac{1}{2}\log\left(1-\hat{q}_{d}+\hat{q}_{0}\right)+\frac{1}{2}\frac{\hat{m}^{2}+\hat{q}_{0}}{1-\hat{q}_{d}+\hat{q}_{0}}
\end{equation}
and for the energetic term
\begin{equation}
    G_{E}(\lambda, t; q_d, q_0, m) = \int D\gamma\int Du^{0}\log\left(\int Du e^{-\frac{\lambda}{2t}\left(g\left(u^{0}\right)-g\left(\sqrt{q_{d}-q_{0}}u+mu^{0}-\sqrt{q_{0}-m^{2}}\gamma\right)\right)^{2}}\right).
\end{equation}
Then one can solve the saddle point equation, which will depend on the choice of the non-linearity $g$, and obtain $\zeta_t(\lambda)$ at the fixed point.

\section{Computation of the KL-Divergence}
\label{appendix:kl}
The Kullback-Leibler (KL) divergence is a type of statistical distance between two probability density functions. Given the two distributions $p_0(x)$, namely the ground-truth distribution of the data, and $p_{t,\mathcal{D}}^{emp}(x)$, namely the empirical distribution of the data according to the model, the full KL divergence between these two functions assumes the following expression
\begin{align}
   \lim_{N\to\infty}\frac{1}{N}\mathbb{E}_{\mathcal{D}}D_{KL}\left[p_0| p_{t,\mathcal{D}}^{emp}\right]
   & =\lim_{N\to\infty}\frac{1}{N}\mathbb{E}_{\mathcal{D}}\left[ \int dx p_{0}(x)\log p_0(x) - \int dx p_{0}(x)\log p_{t,\mathcal{D}}(x)\right]\\ 
   & = -s_0 + \Tilde{D}_{KL}\left[p_0| p_{t}^{emp}\right], 
\end{align}
where $s_0$ is the entropy of the $p_0$ distribution and $\Tilde{D}_{KL}$ is the only time-dependent component of the KL divergence. Since we are studying a data-model where $p_0(x)$ is defined on a support having a lower dimensionality with respect to the $N$-dimensional data-space, we expect the entropy $s_0$ to diverge. This issue might be controlled by adding some noise to either the latent data points $z^{\mu}$ or the features in $F$, but we will not engage into this analysis.  Nevertheless,  for studying the  dependence on $t$ we can  compute the $\Tilde{D}_{KL}$ function in order to find the \textit{generalization time} $t_g$ at which the distance between the two distribution is minimal. We  derive below $\Tilde{D}_{KL}$ in both the linear and non-linear manifold cases by expressing this quantity in terms of time-dependent free-energy function in the REM formalism. 

The time-dependent part of the KL divergence is given by is given by
\begin{equation}
\Tilde{D}_{KL}[p_0|p_t^{emp}] = -\lim_{N \rightarrow \infty}\frac{1}{N}\mathbb{E}_{\mathcal{D}}\int dx\ p_{0}(x)\,\log\,p_{t,\mathcal{D}}^{emp}(x),  
\end{equation}
with $x \sim g(\frac{Fz}{\sqrt{D}})$, $z\sim\mathcal{N}(0,I_D)$, $F\in\mathbb{R}^{N\times D}$, and the empirical score reads
\begin{equation}
    \log p_{t,\mathcal{D}}^{emp}(x) = \log \frac{1}{P\sqrt{2\pi t}^N}\sum_{\mu=1}^P e^{-\frac{1}{2t}\|x-\xi^\mu\|^2} \simeq N\left[\Phi_{t}(x) -\alpha -\frac{1}{2}\log \left(2\pi t\right)\right],
\end{equation}
where
\begin{equation}
    \Phi_{t}(x) = \frac{1}{N} \log \sum_{\mu=1}^P e^{-\frac{1}{2t}\|x-\xi^\mu\|,^2}
\end{equation}
is again minus the free energy density of a REM. For $P, N \to \infty$ with $\alpha = \log P / N$ this concentrates to
\begin{equation}
    \phi(\alpha,t) = \lim_{N\to \infty} \mathbb{E}_{x\sim p_0}[\Phi_{t}(x)],
\end{equation}
and to know this limit we need to compute the large deviation function. 

\subsection{Linear case}

In case of a linear manifold we can compute $\Tilde{D}_{KL}$ in terms of the free-energy of a REM, as in Eq. \eqref{eq:second_piece}. The computation of the free-energy function coincides with the one performed in Appendix \ref{sec:ot_lin}.
\subsection{Non-linear case}

In case of non-linear functions that define the manifold, we are going to employ the replica method to compute the REM free-energy, as we performed for the condensation time. 
The computation coincides with the one performed in Appendix \ref{sec:ot_nlin}.

\end{document}